\documentclass[12pt]{elsarticle}                                          
\usepackage{axodraw4j}
\usepackage{hyperref}
\usepackage{graphicx}                                                         
\usepackage{a41}                                                                
\usepackage{xcolor}                     
\usepackage{slashed}
\usepackage[rflt]{floatflt}
\usepackage{float}
\usepackage{hyperref}
\hypersetup{colorlinks=true, 
linkcolor=blue, filecolor=blue, urlcolor=mygreen}
\usepackage{breakurl} 
\usepackage{times}     %  
\usepackage{array}
\usepackage[english]{babel}
\usepackage[latin1]{inputenc}
\usepackage[T1]{fontenc}
\usepackage{ae}
\usepackage{url}
\usepackage{amsmath, amsthm, amssymb}
\usepackage{slashed}

\usepackage{rotating}
\usepackage{graphicx}
\usepackage{comment}
\newcounter{mmacnt}
\def\restartmma{\setcounter{mmacnt}{0}}
\restartmma \catcode`|=\active
\def|#1|{\mathrm{#1}}
\catcode`|=12
\newenvironment{mma}{
\par\smallskip
\catcode`|=\active
\parskip=0pt\parindent=0pt % locally
\small
\def\In##1\\{%
\def\linebreak{\hfill\break\null\qquad}%
\refstepcounter{mmacnt}
\hangindent=2.5em\hangafter=0
\leavevmode
\llap{\tiny\sffamily In[\arabic{mmacnt}]:=\kern.5em}%
\mathversion{bold}\footnotesiVe$
\displaystyle##1$\normalsiVe
\mathversion{normal}\par
 }%
\def\Print##1\\{%
\def\linebreak{\hfill\break}%
\hangindent=2.5em\hangafter=0
\leavevmode ##1\par}%
\def\Out##1\\{%
\def\linebreak{$\hfill\break\null\hfill$}%
\kern\abovedisplayskip\par
\hangindent=2.5em\hangafter=0
\leavevmode
\llap{\tiny\sffamily Out[\arabic{mmacnt}]=\kern.5em}
\footnotesiVe$\displaystyle##1$
\normalsiVe\hfill\null\par
\kern\belowdisplayskip
}%
\def\Warning##1##2\\{%
\def\linebreak{\hfill\break}%
\hangindent=2.5em\hangafter=0
\leavevmode
{\scriptsiVe##1 : ##2}\par}%
}{%
\par\smallskip
}

\usepackage{color}
\newenvironment{fshaded}{%
\MakeFramed {\FrameRestore}
}%
{\endMakeFramed}

\makeatletter
\def\ps@pprintTitle{%
\let\@oddhead\@empty
\let\@evenhead\@empty
\def\@oddfoot{\reset@font\hfil\thepage\hfil}
\let\@evenfoot\@oddfoot
}
\makeatother
\newcommand{\n}{\nonumber}

\usepackage{tcolorbox}
\usepackage{slashed}
\usepackage{comment}
\begin{document}  %%%%
%%%%%%%%%%%%%%%%%%%%%%
\begin{frontmatter}%%%
%%%%%%%%%%%%%%%%%%%%%%
\title{\large 
\textbf{One-loop formulas for $H\rightarrow 
Z \nu_l\bar{\nu}_l$ 
for $l = e,\mu, \tau$  
in 't Hooft-Veltman gauge}}
\author{Dzung Tri Tran}
% \ead{trantridung@duytan.edu.vn}
\author{Khiem Hong Phan}
\ead{phanhongkhiem@duytan.edu.vn}
\address{\it Institute of Fundamental 
and Applied Sciences, 
Duy Tan University, 
Ho Chi Minh City $700000$, Vietnam\\ 
Faculty of Natural Sciences, Duy Tan University, 
Da Nang City $550000$, Vietnam}
\pagestyle{myheadings}
\markright{}
%%%%%%%%%%%%%%%%%%%%
\begin{abstract} %%%
%%%%%%%%%%%%%%%%%%%%%%%%%%%%%%%%%%%%
In this paper, we present analytical 
results for one-loop contributing to 
the decay processes $H\rightarrow 
Z \nu_l\bar{\nu}_l$ (for $l = 
e, \mu, \tau$). The calculations are 
performed within the Standard Model 
framework in 't Hooft-Veltman gauge.
One-loop 
form factors are then written in terms
of scalar one-loop functions in the 
standard notations of {\tt LoopTools}. 
As a result, one-loop decay rates for
the decay channels can be evaluated 
numerically by using the package.
Furthermore, we analyse the signals
of $H\rightarrow Z \nu_l\bar{\nu}_l$ 
via the production processes 
$e^-e^+ \rightarrow ZH^* \rightarrow Z 
(H^* \rightarrow Z \nu_l\bar{\nu}_l)$
including the initial beam polarizations
at future lepton collider. The Standard 
Model backgrounds such as the processes
$e^-e^+ \rightarrow \nu_l\bar{\nu}_l ZZ$
are also examined in this study. In 
numerical results, 
we find that one-loop corrections
are about $10\%$ contributions 
to the decay rates. They are sizeable 
contributions and should be taken 
into account at future colliders. 
We show that the signals 
$H\rightarrow  Z\nu_l\bar{\nu}_l$
are clearly visible 
at center-of-mass 
energy $\sqrt{s}=250$ GeV and 
are hard to probe at higher-energy 
regions due to the dominant of
the backgrounds.
\end{abstract}
%%%%%%%%%%%%%%%%%%%%%%%%%%%%%%%%%%%%%%%
\end{frontmatter}
%%%%%%%%%%%%%%%%%%%%%%%%%
\section{Introduction}%%%
%%%%%%%%%%%%%%%%%%%%%%%%%%%%%%%%%%%%%%
After discovering 
Standard-Model-like (SM-like) Higgs 
boson at the Large Hadron Collider 
(LHC)~\cite{ATLAS:2012yve,CMS:2012qbp}, 
the high-precision measurements for 
the properties of the SM-like Higgs 
boson are the most important tasks at 
the High-Luminosity 
LHC (HL-LHC)~\cite{Liss:2013hbb,
CMS:2013xfa} and 
future lepton colliders~\cite{Baer:2013cma}. 
In other words, 
all Higgs productions and its decay 
channels should be probed as precisely 
as possible at future colliders.
From these data, we can verify the SM 
at higher-energy regions as well as 
extract the new physics. 
Among the Higgs decay channels, 
$H\rightarrow 
Z \nu_l\bar{\nu}_l$ for $l=e, \mu, \tau$ 
are of interests for 
several aspects. First, one considers 
$Z\rightarrow \nu_l\bar{\nu}_l$ in final
state, the decay processes are 
corresponding to $H\rightarrow$ 
invisible particles which have recently
studied at the LHC~\cite{ATLAS:2022yvh}. 
Search for invisible Higgs-boson 
decays play a key role for 
explaining the existence of 
dark matter. Furthermore, the decay 
channels also contribute to $H\rightarrow$
lepton pair plus missing energy
when $Z\rightarrow$ lepton pair
is concerned in final state. 
These contributions are
also useful to 
evaluate precisely 
the SM backgrounds
for the decay 
rates of $H\rightarrow$ 
lepton pair in final state.  
As above reasons, 
the precise decay rates for 
$H\rightarrow Z \nu_l\bar{\nu}_l$  
could provide an important tool
for testing SM at higher-energy 
scales and for probing new physics.

One-loop contributing to $H\rightarrow 
Z \nu_l\bar{\nu}_l$ have computed 
in~\cite{Chen:2021sgd} and 
for $H\rightarrow 4$ fermions have
presented 
in~\cite{Kniehl:2012rz, Bredenstein:2006rh,
Bredenstein:2006ha}. 
In this paper, we evaluate 
the one-loop contributions for 
the decay processes
$H\rightarrow Z \nu_l\bar{\nu}_l$ for 
$l=e,\mu, \tau$ in 't Hooft-Veltman 
gauge. In comparison 
with the previous calculations, we perform 
this computation with the following
advantages. First, we focus on the 
analytical calculations 
for the decay channels
and show a clear 
analytical structure for the one-loop
amplitude of 
$H\rightarrow Z \nu_l\bar{\nu}_l$ 
in this paper.
As a result, we can explain and 
extract the dominant
contributions to the decay widths when these
are necessary
(the dominant contributions 
are from $Z$-pole diagrams, or 
the diagrams of $H\rightarrow Z Z^*
\rightarrow Z \nu_l\bar{\nu}_l$ in the
decay channels as we show in later sections). 
Furthermore, 
off-shell Higgs decays are also 
valid in our work.
In addition, one can generalize
the couplings of Nambu-Goldstone bosons
with Higgs, gauge bosons, etc 
(as our previous work in~\cite{Phan:2021pcc}).
We are easily to extend our results 
to many of beyond the Standard models.  
Since Nambu-Goldstone bosons 
play the same role like the 
changed Higgs in the 
extensions of the Standard
Model Higgs sector.
Last but not least, the
signals of $H\rightarrow 
Z \nu_l\bar{\nu}_l$ via Higgs 
productions at future 
lepton colliders are studied in 
our works.
In further detail, one-loop form factors 
are expressed in terms of scalar 
one-loop Passarino-Veltman functions
(called as PV-functions hereafter)
in the standard notations 
of {\tt LoopTools}. As a result, 
one can evaluate the decay rates 
numerically by using the package. 
Moreover, 
the signals of $H\rightarrow 
Z \nu_l\bar{\nu}_l$ through Higgs 
productions at future lepton collider, 
for instance, the processes 
$e^-e^+ \rightarrow ZH^* \rightarrow Z 
(Z \nu_l\bar{\nu}_l)$
with including initial beam 
polarizations are generated. 
The Standard Model 
backgrounds such as
$e^-e^+ \rightarrow \nu_l\bar{\nu}_l ZZ$
are also included in this
analyse. In phenomenological results, 
we find that one-loop corrections
are about $10\%$ contributions 
to the decay rates. They are 
sizeable contributions and 
should be taken into account 
at future colliders. 
We show that the signals 
$H\rightarrow  Z\nu_l\bar{\nu}_l$
are clearly visible at center-of-mass 
energy $\sqrt{s}=250$ GeV and 
these are hard
to probe at higher-energy 
regions due to the dominant of
the backgrounds.

The layout of the paper is 
as follows: 
In section $2$, we present 
the calculations for  $H\rightarrow 
Z \nu_l\bar{\nu}_l$ in detail. 
We then show phenomenological 
results for the computations. 
Decay rates for on-shell and 
off-shell Higgs decay modes 
are studied with including 
the unpolarized and 
longitudinally polarized 
$Z$ boson 
in the final states.
The signals of $H\rightarrow 
Z \nu_l\bar{\nu}_l$ via 
the Higgs productions at 
future lepton colliders are also 
generated in this section. 
Conclusions for this work
are discussed in the section $4$. 
In the appendies, we first summary 
all tensor reduction formulas for
one-loop integrals appear in this 
work. Numerical checks for 
the calculations
are 
presented. All self-energy and 
counter-terms for the decay 
processes are shown in detail. 
One-loop Feynman diagrams in 
't Hooft-Veltman gauge for this 
decay channels are 
shown in the appendix $E$.
%%%%%%%%%%%%%%%%%%%%%%%%%
\section{Calculations}%%%
%%%%%%%%%%%%%%%%%%%%%%%%%
We are going to present
the calculations for $H(p_H)\rightarrow 
Z(q_1) \nu_l(q_2)\bar{\nu}_l(q_3)$
in detail. For these computations, 
we are working in 
t 'Hooft-Veltman gauge. Within the SM 
framework, all Feynman diagrams can 
be grouped in
several classifications showing 
in the Appendix $E$. In group $G_0$, 
we have tree Feynman diagram contributing
to the decay processes. For group $G_1$,
we include all one-loop Feynman 
diagrams correcting to the vertex 
$Z\nu_l\bar{\nu}_l$. We then list 
all $Z$-pole Feynman diagrams in 
group $G_2$ and non $Z$-pole diagrams 
in group $G_3$. The counterterm 
diagrams for this decay channels are 
classified into group $G_4$.  

In general, the amplitude 
for $H(p_H)\rightarrow 
Z(q_1) \nu_l(q_2)\bar{\nu}_l(q_3)$
can be decomposed by 
the following Lorentz structure: 
\begin{eqnarray}
\label{ampHZvv}
 \mathcal{A}_{H\rightarrow 
Z \nu_l\bar{\nu}_l}
&=&
\Big\{ 
F_{00}
g^{\mu \nu}
+
F_{12}\,
q_1^\nu q_2^\mu
+
F_{13}
\,
q_1^\nu q_3^\mu 
\Big\}
\Big[
\bar{u} (q_2)
\gamma_\nu
P_L
v (q_3)
\Big]
\varepsilon_\mu^* (q_1).
\end{eqnarray}
Where $F_{00}$,  $F_{12}$ and 
$F_{13}$ are form factors 
including both tree-level and 
one-loop diagram contributions. 
The form factors are functions 
of the Mandelstam invariants 
such as
$s_{ij} = (q_i + q_j)^2$ for 
$i \neq j = 1,2,3$ and 
mass-squared in one-loop diagrams. 
One also verifies
that $s_{12} + s_{13} + s_{23} 
= M_H^2 + M_Z^2$. In Eq.~(\ref{ampHZvv}), 
projection operator
$P_L= (1-\gamma_5)/2$ is taken into account
and
the term $\varepsilon_\mu^* (q_1)$ is 
polarization vector of final $Z$ boson. 
Our computations can be summarized 
as follows. We first write 
down Feynman amplitude 
for all diagrams mentioned above. 
By using {\tt Package-X}~\cite{Patel:2015tea},  
all Dirac traces and Lorentz 
contractions in $d$ dimensions are performed. 
The amplitudes
are then casted into tensor 
one-loop integrals. The tensor 
integrals are next reduced to 
scalar PV-functions
~\cite{Denner:2005nn}. It is noted 
that all the relevant tensor 
reduction formulas 
are shown in appendix $A$. The 
PV-functions can be evaluated 
numerically 
by using  {\tt LoopTools}.

All form factors
are calculated from Feynman diagrams
in 't Hooft-Veltman gauge and their 
expressions are presented in this 
section.   
For tree-level diagram, the 
form factor is given by:
\begin{eqnarray}
F_{00}^{(G_0)} (s_{12},s_{13},s_{23})
&=& 
\dfrac{2 \pi \alpha}{s_W^2 c_W^3}
\dfrac{M_W}{s_{23}-M_Z^2 
+ i \Gamma_Z M_Z}. 
%%%%%%%%%%%%%%%%%%%%%%%%%%%%%%%%%%%
\end{eqnarray}
Where $s_W (c_W)$ is sine (cosine) 
of Weinberg angle, respectively and 
$\Gamma_Z$ is decay width of $Z$ 
boson. 

At one-loop level, all form factors
are taken the form of
\begin{eqnarray}
F_{ij} &=&\sum\limits_{G=\{G_1, 
\cdots, G_4\}}F_{ij}^{(G)} (s_{12},s_{13},s_{23}), 
\quad \textrm{for}\; 
ij = \{00, 12, 13\}.
\end{eqnarray}
Where $\{G_1, \cdots, G_4\} = 
\{\text{group 1},  
\cdots, \text{group 4}\}$
are corresponding to the groups 
of Feynman diagrams in the appendix 
$E$. By considering each 
group of Feynman diagram, 
analytic results for all form 
factors are presented in the 
following paragraphs. 
Taking the attribution 
from group $G_1$, we have 
one-loop form factors 
accordingly 
%%%%%%%%%%%%%%%%%%%%%%%%%%%%%%%%%%%
\begin{eqnarray}
F_{00}^{(G_1)} (s_{12},s_{13},s_{23})
&=& -\dfrac{\alpha^2 }{8 s_W^4 c_W^5}
\dfrac{M_W}{s_{23}-M_Z^2 
+ i \Gamma_Z M_Z}
\times
\n \\
&&\times
\Bigg\{
-8 c_W^4 
B_0(s_{23},M_W^2,M_W^2)
-2 \Big[c_W^2 (4 s_W^2-2)
+1\Big] B_0(s_{23},0,0)
\n \\
&&
-8 c_W^4 
\Big[2 C_{00}
-s_{23} (C_1
+C_2)\Big](0,s_{23},0,0,M_W^2,M_W^2)
\\
&&
-4 c_W^2 (2 s_W^2-1) 
\Big[
M_W^2 C_0
+s_{23} C_2
-2 C_{00}
\Big] (s_{23},0,0,0,0,M_W^2)
\n \\
&&
+ \Big[
4 C_{00}
- 2 s_{23} C_2
-2 M_Z^2 C_0
\Big] (s_{23},0,0,0,0,M_Z^2)
+(2 c_W^2 + 1)
\Bigg\}, \n \\
F_{ij}^{(G_1)} (s_{12},s_{13},s_{23})
&=& 0, 
\quad \text{for}\quad ij =\{12, 13\}.
\end{eqnarray}
%%%%%%%%%%%%%%%%%%%%%%%%%%%%%%%%%%%
For group $G_2$ of Feynman diagram, 
the form factors can be divided into
the fermion and 
boson parts as follows:
\begin{eqnarray}
%%%%%%%%%%%%%%%%%%%%%%%%%%%%%%%%%%%
F_{00}^{(G_2)} (s_{12},s_{13},s_{23})
&=& \dfrac{\alpha^2}{24\; s_W^4 c_W^7 M_W}
\dfrac{1}{(s_{23}-M_Z^2 + i \Gamma_Z M_Z)^2}
\Big[
\sum\limits_{f} N_f^{C} 
F_{00, f}^{(G_2)} 
+ F_{00, b}^{(G_2)} 
\Big].
\end{eqnarray}
Where $N_f^C$ is color number. It takes
$3$ for quarks and $1$ for leptons. 
For the fermion contributions, we take 
top quark loop as example, analytic
results are written as
\begin{eqnarray}
%%%%%%%%%%%%%%%%%%%%%%%%%%%%%%%%%%%
F_{00, f}^{(G_2)} 
&=&  
2 c_W^2 M_W^2 
\Big[
8 s_W^2 (4 s_W^2-3)+9
\Big] \times 
\\
&&
\hspace{3cm}
\times 
\Big[
A_{0}(m_t^2)
+ s_{23} B_{1}(s_{23},m_t^2,m_t^2)
- 2 B_{00}(s_{23},m_t^2,m_t^2) 
\Big]
\n \\
&&\hspace{0cm}
+2 m_t^2 c_W^2 
\Big\{
9 M_W^2
-2 c_W^2 (M_Z^2-s_{23}) 
\Big[4 s_W^2 (4 s_W^2-3)+9\Big]
\Big\} 
B_{0}(s_{23},m_t^2,m_t^2)
\n \\
&&\hspace{0cm}
+ m_t^2 c_W^4 (s_{23}-M_Z^2)
\Big\{
36 m_t^2+\Big[8 s_W^2 (4 s_W^2-3)
+9\Big] (s_{12}+s_{13})
\Big\} \times
\nonumber\\
&&
\hspace{7cm}
\times 
C_{0}(M_Z^2,s_{23},M_H^2,
m_t^2,m_t^2,m_t^2)
\n \\
&&\hspace{0cm}
-m_t^2 c_W^4 (M_Z^2-s_{23}) 
\Big\{
(M_H^2+5 M_Z^2-s_{23}) 
\Big[8 s_W^2 (4 s_W^2-3)+9\Big]
\n \\
&&\hspace{3cm}
+8 s_W^2 (3-4 s_W^2) 
(s_{12}+s_{13})
\Big\} 
C_{1}(M_Z^2,s_{23},M_H^2,
m_t^2,m_t^2,m_t^2)
\n \\
&&\hspace{0cm}
-2 m_t^2 c_W^4 (M_Z^2-s_{23}) 
\Big\{
9 M_H^2+
\Big[8 s_W^2 (4 s_W^2-3)
+9\Big] (s_{12}+s_{13})
\Big\} 
\times 
\nonumber\\
&&
\hspace{7cm}
\times 
C_{2}(M_Z^2,s_{23},M_H^2,m_t^2,m_t^2,m_t^2)
\n \\
&&\hspace{0cm}
+8 m_t^2 c_W^4 (M_Z^2-s_{23}) 
\Big[8 s_W^2 (4 s_W^2-3)+9\Big] 
C_{00}(M_Z^2,s_{23},M_H^2,m_t^2,m_t^2,m_t^2).
\nonumber
\end{eqnarray}
The contribution from boson part 
reads
\begin{eqnarray}
F_{00, b}^{(G_2)} 
&=& 8 c_W^6 M_W^2 s_{23}
-6 M_W^2 c_W^2 
\Big[3 c_W^2 (4 c_W^2-1)+s_W^2\Big] 
A_{0}(M_W^2) -3 M_W^2 c_W^2 
\Big[
A_{0}(M_Z^2) + A_{0}(M_H^2)
\Big]
\n \\
&&\hspace{0cm}
+\dfrac{3}{2} c_W^4 M_H^2 (M_Z^2-s_{23}) 
B_{0}(M_H^2,M_Z^2,M_Z^2)
+ 12 M_W^2 s_W^4 c_W^4 (M_Z^2-s_{23}) 
B_{0}(M_Z^2,M_W^2,M_W^2)
\n \\
&&\hspace{0.0cm}
+3 c_W^4 (M_Z^2-s_{23}) 
\Big[
c_W^4 (M_H^2+24 M_W^2)
-2 M_H^2 s_W^2 c_W^2
+M_H^2 s_W^4
\Big] 
B_{0}(M_H^2,M_W^2,M_W^2)
\n \\
&&\hspace{0.0cm}
+12 M_W^2 c_W^2 
\Big\{
(2 M_W^2 + 5 s_{23}) c_W^4
-2 M_W^2 s_W^4
\\
&&\hspace{4cm}
+(M_Z^2-s_{23}) 
\Big[s_W^4-2 c_W^2 (c_W^2+1)\Big] c_W^2
\Big\} 
B_{0}(s_{23},M_W^2,M_W^2)
\n \\
&&\hspace{0.0cm}
+6 M_W^2 c_W^2 (M_Z^2-s_{23}) 
\Big[B_{0}(M_Z^2,M_H^2,M_Z^2)
+B_{0}(s_{23},M_H^2,M_Z^2)
\Big]
-12 M_W^4 B_{0}(s_{23},M_H^2,M_Z^2)
\n \\
&&\hspace{0.0cm}
-4 M_W^2 c_W^2 \Big[40 s_W^2 (4 s_W^2-3)+63\Big] 
B_{00}(s_{23},0,0)
+12 M_W^2 c_W^2 B_{00}(s_{23},M_H^2,M_Z^2)
\n \\
&&\hspace{0.0cm}
+12 M_W^2 c_W^2 
\Big(9 c_W^4-2 s_W^2 c_W^2+s_W^4\Big) 
B_{00}(s_{23},M_W^2,M_W^2)
\n \\
&&\hspace{0.0cm}
+
\dfrac{9}{2} c_W^4 M_H^2 (M_Z^2-s_{23}) 
B_{0}(M_H^2,M_H^2,M_H^2)
\n \\
&&\hspace{0.0cm}
+2 M_W^2 c_W^2 s_{23} 
\Big[
40 s_W^2 (4 s_W^2-3)+63
\Big] 
B_{1}(s_{23},0,0)
+
24 c_W^6 M_W^2 s_{23} 
B_{1}(s_{23},M_W^2,M_W^2)
\n \\
&&\hspace{0.0cm}
-12 M_W^2 c_W^6 (M_Z^2-s_{23}) 
% \Big[4 M_Z^2 s_W^2
% +c_W^2 (4 M_Z^2+s_{12}+s_{13})
% \Big] 
[
4 M_Z^2+
c_W^2 (s_{12}+s_{13})
]
C_{1}(M_Z^2,s_{23},M_H^2,M_W^2,M_W^2,M_W^2)
\n \\
&&\hspace{0.0cm}
+24 M_W^2 c_W^6 (M_Z^2-s_{23}) 
% \Big[c_W^2 (M_Z^2-s_{23})
% -(c_W^2+1) (s_{12}+s_{13})
% \Big] 
(-c_W^2 M_H^2-s_{12}-s_{13})
C_{2}(M_Z^2,s_{23},M_H^2,M_W^2,M_W^2,M_W^2)
\n \\
&&\hspace{0.0cm}
-12 M_W^2 c_W^4 (M_Z^2-s_{23}) 
\Big\{
2 c_W^4 \Big[
2 M_W^2+5 M_Z^2
-2 (s_{12}+s_{13})
\Big]
-(M_H^2+2 M_W^2) s_W^4
\n \\
&&\hspace{2.5cm}
+s_W^2 c_W^2 
\Big(M_H^2+2 M_W^2+s_{12}+s_{13}\Big)
\Big\} 
C_{0}(M_Z^2,s_{23},M_H^2,M_W^2,M_W^2,M_W^2)
\n \\
&&\hspace{0.0cm}
-12 c_W^4 (M_Z^2-s_{23}) 
\Big[
s_W^2 (s_W^2-2 c_W^2) (M_H^2+2 M_W^2)
\n \\
&&\hspace{5.0cm}
+ (M_H^2+18 M_W^2) c_W^4
\Big] 
C_{00}(M_Z^2,s_{23},M_H^2,M_W^2,M_W^2,M_W^2)
\n \\
&&\hspace{0cm}
% % \textcolor{red}{
% % %%%%
% % + 3 M_H^2 c_W^2 
% % \Big[
% % C_{0}
% % +6 c_W^2 (s_{23}-M_Z^2) 
% % C_{00}
% % \Big]
% % (M_H^2,s_{23},M_Z^2,M_H^2,M_H^2,M_Z^2)
% % %%%
% % }
 \n \\
&&\hspace{0cm}
+ 18 M_H^2 c_W^2 (s_{23}-M_Z^2) 
\Big[
-M_W^2 C_{0}
+ c_W^2 
C_{00}
\Big]
(M_H^2,s_{23},M_Z^2,M_H^2,M_H^2,M_Z^2)
\n \\
&&\hspace{0cm}
+ (M_Z^2-s_{23}) \Big[
12 M_W^4 C_{0}
-6 c_W^2 (M_H^2 c_W^2+2 M_W^2) 
C_{00}
\Big] 
(M_Z^2,M_H^2,s_{23},M_H^2,M_Z^2,M_Z^2)
.\n
\end{eqnarray}
Other one-loop 
form factors are given the 
same convention as
\begin{eqnarray}
F_{12}^{(G_2)} (s_{12},s_{13},s_{23})
&=& F_{13}^{(G_2)} (s_{13},s_{12},s_{23})
\\
&=&
- \dfrac{\alpha^2}{12 s_W^4 c_W^5 M_W}
\dfrac{1}{s_{23}-M_Z^2 + i \Gamma_Z M_Z}
\Big[
\sum\limits_{f} N_f^{C}
F_{12, f}^{(G_2)} 
+
F_{12, b}^{(G_2)}
\Big]. 
\nonumber
\end{eqnarray}
Each part in the above equation
reads the form of 
(we also take top quark 
loop as an example for fermion 
contributions)
\begin{eqnarray}
F_{12,f}^{(G_2)} 
&=&
9 m_t^2 c_W^2 
C_{1}(M_Z^2,s_{23},M_H^2,m_t^2,m_t^2,m_t^2)
\n \\
&&
+ m_t^2 c_W^2 
\Big[8 s_W^2 (4 s_W^2-3)+9\Big]
\Big[
C_{0}
+
4
\Big(
C_{2}
+
C_{12}
+
C_{22}
\Big)
\Big] 
(M_Z^2,s_{23},M_H^2,m_t^2,m_t^2,m_t^2)
\nonumber\\
\end{eqnarray}
and 
\begin{eqnarray}
F_{12,b}^{(G_2)} 
&=&
-12 c_W^2 M_W^2 
\Big[
\big(5 c_W^4-2 c_W^2 s_W^2+s_W^4\big) 
C_{0}
+
C_{1}
\Big] (M_Z^2,s_{23},M_H^2,M_W^2,M_W^2,M_W^2)
\n \\
&&
+3 c_W^2 M_H^2 
\Big[
C_{12}(M_Z^2,M_H^2,s_{23},M_H^2,M_Z^2,M_Z^2)
\\
&&\hspace{4.0cm}
-3 
\Big(
C_{1}
+
C_{11}
+
C_{12}
\Big) (M_H^2,s_{23},M_Z^2,M_H^2,M_H^2,M_Z^2)
\Big]
\n \\
&&
+6 M_W^2 
\Big[
C_{1}
+
C_{2}
+
C_{12}
\Big] 
(M_Z^2,M_H^2,s_{23},M_H^2,M_Z^2,M_Z^2)
\n \\
&&
-6 c_W^2 
\Big[
s_W^2 (s_W^2 - 2 c_W^2) (M_H^2+2 M_W^2)
+ c_W^4 (M_H^2+18 M_W^2)
\Big] 
\n \\
&&\hspace{4.0cm}
\times
\Big[
C_{2}
+
C_{12}
+
C_{22}
\Big] 
(M_Z^2,s_{23},M_H^2,M_W^2,M_W^2,M_W^2)
. \n
\end{eqnarray}
We change to the contributions of 
all Feynman diagrams in group $G_3$. 
For this group, there is no $Z$-pole 
diagrams including in one-loop form
factors. But we have one-loop box
diagrams. There are triple gauge 
boson vertex
and the propagator of lepton, or 
two propagators of leptons
in one-loop box diagrams, we hence 
have tensor box integrals which 
the highest rank
is up to $R=2$
in the amplitude. It is explained
that the corresponding form 
factors are expressed in terms 
of the PV-functions $C$- 
and up to $D_{33}$-coefficients.
\begin{eqnarray}
%%%%%%%%%%%%%%%%%%%%%%%%%%%%%%%%%%%
F_{00}^{(G_3)}
&=& 
\dfrac{\alpha^2 M_W}{4 s_W^4 c_W^5}
\times  \\
&&\hspace{0cm} \times
\Bigg\{
2 c_W^4 
\Big[
(2 s_W^2-1) C_{0}(M_Z^2,0,s_{13},0,0,M_W^2)
+ (3 c_W^2+1) C_{0}(0,s_{23},0,0,M_W^2,M_W^2)
\n \\
&&\hspace{3.5cm}
% \textcolor{red}{
+ C_{2}(0,M_H^2,s_{13},0,M_W^2,M_W^2)
+ C_{2}(0,M_H^2,s_{12},0,M_W^2,M_W^2)
\Big]
% }
% \n \\
% &&\hspace{3cm}
% \textcolor{blue}{
% +2 c_W^4 C_{2}(0,M_H^2,s_{13},0,M_W^2,M_W^2)
% -2 c_W^4 C_{2}(0,M_H^2,s_{12},0,M_W^2,M_W^2)
% \Big]
% }
\n \\
&&\hspace{0cm}
+C_{0}(M_Z^2,0,s_{13},0,0,M_Z^2)
+C_{2}(0,M_H^2,s_{13},0,M_Z^2,M_Z^2)
+C_{2}(0,M_H^2,s_{12},0,M_Z^2,M_Z^2)
\n \\
&&\hspace{0cm}
+8 c_W^6 
\Big[
D_{00}(s_{12},M_Z^2,s_{23},0,0,
M_H^2,0,M_W^2,M_W^2,M_W^2)
\n \\
&&\hspace{6.0cm}
+ D_{00}(s_{13},M_Z^2,s_{23},0,0,
M_H^2,0,M_W^2,M_W^2,M_W^2)
\Big]
\n \\
&&\hspace{0cm}
- 2 c_W^6 
\Big[
(2 M_Z^2+s_{12}) 
D_{1}(s_{12},M_Z^2,s_{23},0,0,
M_H^2,0,M_W^2,M_W^2,M_W^2)
\n \\
&&\hspace{4.0cm}
+ (2 M_Z^2+s_{13}) 
D_{1}(s_{13},M_Z^2,s_{23},0,0,
M_H^2,0,M_W^2,M_W^2,M_W^2)
\Big]
\n \\
&&\hspace{0cm}
+ c_W^4 
\Big\{
\Big[2 c_W^2 (3 M_H^2-2 s_{23}-3 s_{13})
+s_W^2 (s_{13}-M_H^2)
\Big] 
\times
\n \\
&&\hspace{6cm} \times
D_{3}(s_{13},M_Z^2,s_{23},
0,0,M_H^2,0,M_W^2,M_W^2,M_W^2)
\n  \\
&&\hspace{1cm}
+ \Big[
s_W^2 (s_{12} - M_H^2)
+ 2 c_W^2 (3 M_H^2 
- 2 s_{23} - 3 s_{12})
\Big] 
\times
\n \\
&&\hspace{6cm} \times
D_{3}(s_{12},M_Z^2,s_{23},0,0,
M_H^2,0,M_W^2,M_W^2,M_W^2)
\Big\}
\n \\
&&\hspace{0cm}
+c_W^4 (3 c_W^2+1) 
\Big[
(M_W^2-s_{12}) 
D_{0}(s_{12},M_Z^2,s_{23},0,0,
M_H^2,0,M_W^2,M_W^2,M_W^2)
\n \\
&&\hspace{4.0cm}
+ (M_W^2-s_{13}) 
D_{0}(s_{13},M_Z^2,s_{23},0,0,
M_H^2,0,M_W^2,M_W^2,M_W^2)
\Big]
\n \\
&&\hspace{0cm}
+(s_{12}-M_Z^2) 
\Big[2 c_W^4 (1-2 s_W^2) 
D_{2}(M_Z^2,s_{12},M_H^2,
s_{13},0,0,0,0,M_W^2,M_W^2)
\n \\
&&\hspace{6.0cm}
-D_{2}(M_Z^2,s_{12},M_H^2,
s_{13},0,0,0,0,M_Z^2,M_Z^2)
\Big]
\n \\
&&\hspace{0cm}
+s_W^2 c_W^4 
\Big[
(M_Z^2-s_{13}) 
D_{2}(s_{13},M_Z^2,s_{23},0,0,
M_H^2,0,M_W^2,M_W^2,M_W^2)
\n \\
&&\hspace{4.0cm}
+ (M_Z^2 - s_{12}) 
D_{2}(s_{12},M_Z^2,s_{23},0,0,
M_H^2,0,M_W^2,M_W^2,M_W^2)
\Big]
\n \\
&&\hspace{0cm}
+(s_{23}+s_{12}) 
\Big[
2 c_W^4 (2 s_W^2-1) 
D_{3}(M_Z^2,s_{12},M_H^2,
s_{13},0,0,0,0,M_W^2,M_W^2)
\n \\
&&\hspace{6.0cm}
+D_{3}(M_Z^2,s_{12},M_H^2,
s_{13},0,0,0,0,M_Z^2,M_Z^2)
\Big]
\n \\
&&\hspace{0cm}
+ \Big[M_Z^2 D_{0} 
+ s_{12} D_{1} -2 D_{00}\Big]
(M_Z^2,s_{12},M_H^2,
s_{13},0,0,0,0,M_Z^2,M_Z^2)
\n \\
&&\hspace{0cm}
+ 2 c_W^4 (1-2 s_W^2) 
\Big[
2 D_{00}
- s_{12} D_{1}
- M_W^2 D_{0}
\Big]
(M_Z^2,s_{12},M_H^2,
s_{13},0,0,0,0,M_W^2,M_W^2)
\Bigg\}
. \n
\end{eqnarray}
In addition, we have other form
factors which are expressed as 
follows:
\begin{eqnarray}
F_{12}^{(G_3)} 
&=& 
\dfrac{\alpha^2 M_W}{2 s_W^4 c_W^5}
\Bigg\{
\Big[
D_{2}
+D_{12}
+D_{23}
\Big] 
(M_Z^2,s_{12},M_H^2,
s_{13},0,0,0,0,M_Z^2,M_Z^2)
\n \\
&&\hspace{1.5cm}
-4 c_W^6 
\Big[
D_{3}
+
D_{13}
\Big](s_{13},M_Z^2,
s_{23},0,0,M_H^2,0,M_W^2,M_W^2,M_W^2)
\\
&&\hspace{1.5cm}
-2 c_W^4 (1- 2 s_W^2) 
\Big[ 
D_{2}
+
D_{12}
+
D_{23}
\Big]
(M_Z^2,s_{12},M_H^2,s_{13},
0,0,0,0,M_W^2,M_W^2)
\n \\
&&\hspace{0cm}
+2 c_W^4 
\Big[
2 c_W^2 
\Big(
D_{11}
+
D_{12}
\Big)
+(s_W^2 - c_W^2) D_{2}
\Big]
(s_{12},M_Z^2,s_{23},0,0,
M_H^2,0,M_W^2,M_W^2,M_W^2)
\Bigg\}
, \n \\
%%%%%%%%%%%%%%%%%%%%%%%%%%%%%%%%%%
\label{rep1}
F_{13}^{(G_3)} 
&=& 
\dfrac{\alpha^2 M_W}{2 s_W^4 c_W^5}
\Bigg\{
-\Big[
D_{13}
+
D_{33}
\Big] (M_Z^2,s_{12},M_H^2,
s_{13},0,0,0,0,M_Z^2,M_Z^2)
\n \\
&&\hspace{1.5cm}
-4 c_W^6 
\Big[
D_{3}
+
D_{13}
\Big] (s_{12},M_Z^2,s_{23},0,0,
M_H^2,0,M_W^2,M_W^2,M_W^2)
\\
&&\hspace{1.5cm}
+2 c_W^4 (1-2 s_W^2) 
\Big[
D_{13}
+
D_{33}
\Big] (M_Z^2,s_{12},M_H^2,s_{13},
0,0,0,0,M_W^2,M_W^2)
\n \\
&&\hspace{0cm}
+ 2 c_W^4
\Big[
2 c_W^2 
\Big(
D_{11}
+
D_{12}
\Big)
+ \big(s_W^2 - c_W^2\big)
D_{2}
\Big] (s_{13},M_Z^2,s_{23},0,0,
M_H^2,0,M_W^2,M_W^2,M_W^2)
\Bigg\}
. \n
\end{eqnarray}
It is stress that one has the 
following relation:
\begin{eqnarray}
\label{rep2}
 F_{12}^{(G_3)}(s_{12},s_{13},s_{23})
 = F_{13}^{(G_3)}(s_{13},s_{12},s_{23}).
\end{eqnarray}
If we apply several transformations
for box-functions, we can confirm 
the relation. The transformations
for box-functions are not presented 
in this subsection. 
Instead of this, we verify the 
relation by numerical check. 
One finds 
that two representations for 
$F_{13}^{(G_3)}$
in (\ref{rep1}) and (\ref{rep2}) 
are good
agreement up to last digit at several
sampling points.

Having all form 
factors, the decay rates 
can be evaluated
as follows:
\begin{eqnarray}
\Gamma_{H\rightarrow Z\nu_l \bar{\nu_l}} &=& 
\dfrac{1}{256 \pi^3 M_H^3 M_Z^2}
\int \limits_{4 m_{\nu_l}^2}^{(M_H-M_Z)^2} d s_{23}
\int 
\limits_{s_{12}^{\text{min}}}^{s_{12}^{\text{max}}}
ds_{12}
\times
\n \\
&&\times
\; 
\Bigg\{
\Big(M_Z^2 (2 s_{23}-M_H^2) 
+ s_{12} s_{13}\Big)  \Big[
\Big|F_{00}^{(G_0)} \Big|^2
+ 
2\,\mathcal{R}\text{e}
\Big(
F_{00}^{(G_0),*}\cdot
\sum_{i=1}^4
F_{00}^{(G_i)} \Big)
\Big]
\\
&&
+ 
\Big(M_H^2 M_Z^2 - s_{12}
s_{13}\Big)
\Big[
\Big(M_Z^2-s_{12}\Big)
\mathcal{R}\text{e}
\Big(
F_{00}^{(G_0),*}\cdot \sum_{i=1}^4
F_{12}^{(G_i)} 
\Big)
+
\big( s_{12} 
\leftrightarrow 
s_{13} \big)
\Big]
\Bigg\}
. \n
\end{eqnarray}
Where 
\begin{eqnarray}
\label{sminmax}
s_{12}^{\text{max},\text{min}}
=
\dfrac{1}{2}
\Bigg\{
M_H^2 + M_Z^2 - s_{23}
\pm
\sqrt{
\Big(M_H^2 + M_Z^2 - s_{23}\Big)^2 
- 4 M_H^2 M_Z^2 
}
\Bigg\}
.
\end{eqnarray}
The polarized $Z$ boson case 
is next considered. The longitudinal
polarization vectors 
for $Z$ bosons are defined in the 
rest frame of Higgs boson: 
\begin{eqnarray}
 \varepsilon_{\mu}(q_1, \lambda=0) 
 = \dfrac{4 M_{H^*}^2 \; q_{1, \mu}
 - (s_{23}+M_Z^2)\; p_{H,\mu} }
 {M_Z 
 \sqrt{\lambda(s_{23},4M_{H^*}^2,M_Z^2)
 }}.
\end{eqnarray}
Where off-shell Higgs mass is 
given by
$p_H^2=M_{H^*}^2 \neq M_H^2$. 
The Kall\"en function is defined as 
$\lambda (x, y, z) = 
(x - y - z)^2 - 4 yz$.
We then arrive at
\begin{eqnarray}
\Gamma_{H\rightarrow 
Z_L \nu_l \bar{\nu_l}} 
&=& 
\dfrac{1}{256 \pi^3 M_{H^*}^3 M_Z^2}
\int \limits_{4 m_{\nu_l}^2}^{(M_{H^*}-M_Z)^2} 
d s_{23}
\int \limits_{s_{12}^{\text{min}}}
^{s_{12}^{\text{max}}} ds_{12}
\dfrac{
\Big(s_{23}-4 M_{H^*}^2+M_Z^2\Big) 
\Big(s_{12} s_{13}-M_Z^2 M_{H^*}^2\Big)
}
{[s_{23}-M_Z^2 -4 M_{H^*}^2]^2- 16 M_Z^2 M_{H^*}^2 }
% {\lambda \Big(4 M_H^2,M_Z^2,s_{23}\Big)}
\times
\n \\
&&\times
\; 
\Bigg\{
\Big(s_{23}-4 M_{H^*}^2+M_Z^2\Big)
\Big[ \Big|F_{00}^{(G_0)}\Big|^2 
+
2\, \mathcal{R}\text{e}
\Big(
F_{00}^{(G_0),*} 
\times
\sum_{i=1}^4
F_{00}^{(G_i)}
\Big)
\Big]
\\
&&
\hspace{0.75cm}
+
\Big[
\Big(
s_{23}^2
- M_Z^4
+ 4 M_{H^*}^2 M_Z^2
\Big)
+ 
\Big(
s_{23}
-4 M_{H^*}^2
+ M_Z^2 \Big)s_{12}
\Big]
\times
\n \\
&&\hspace{6.6cm} \times
\mathcal{R}\text{e}
\Big(
F_{00}^{(G_0),*} \sum_{i=1}^4 
F_{12}^{(G_i)} 
\Big)
+
\big(s_{12} 
\leftrightarrow 
s_{13} \big)
\Bigg\}
. \n
\end{eqnarray}
Where $s_{12}^{\text{min, max}}$
are obtained as in equation 
(\ref{sminmax}) in 
which $M_H$ is replaced
by off-shell Higgs mass $M_{H^*}$.

In the next section, we show 
phenomenological results for the
decay processes. Before generating
the data, numerical checks for 
the calculations are performed. 
The $UV$-
finiteness and $\mu^2$-independent of 
the results 
are verified. Numerical results 
for this check are shown in Appendix $B$.
One finds the results are good stability
over $14$ digits. 
%%%%%%%%%%%%%%%%%%%%%%%%%%%%%%%%%%%%%%%%
\section{Phenomenological results}   %%%
%%%%%%%%%%%%%%%%%%%%%%%%%%%%%%%%%%%%%%%%
In the phenomenological results, 
we use the following input parameters: 
$M_Z = 91.1876$ GeV, 
$\Gamma_Z  = 2.4952$ GeV, 
$M_W = 80.379$ GeV, $\Gamma_W  = 2.085$ GeV, 
$M_H =125.1$ GeV, $\Gamma_H =4.07\cdot 10^{-3}$ GeV. 
The lepton masses are given: $m_e =0.00052$ GeV,
$m_{\mu}=0.10566$ GeV and $m_{\tau} = 1.77686$ GeV.
For quark masses, one takes $m_u= 0.00216$ GeV
$m_d= 0.0048$ GeV, $m_c=1.27$ GeV, $m_s = 0.93$ GeV, 
$m_t= 173.0$ GeV, and $m_b= 4.18$ GeV. 
We work in the so-called $G_{\mu}$-scheme
in which the Fermi constant 
is taken $G_{\mu}=1.16638\cdot 
10^{-5}$ GeV$^{-2}$ and the 
electroweak coupling can be calculated 
appropriately
as follows:
\begin{eqnarray}
\alpha = \sqrt{2}/\pi G_{\mu}
M_W^2(1-M_W^2/M_Z^2)
= 1/132.184.
\end{eqnarray}

We then present the phenomenological 
results in the following subsections. 
We first mention about the decay rates
for on-shell Higgs decay 
$H\rightarrow Z \nu_l\bar{\nu}_l$.
In the 
Table~\ref{DRON}, the decay rates
for on-shell Higgs decay to  
$Z \nu_e\bar{\nu}_e$ are generated. 
In the first column, the cuts for 
invariant mass of final neutrino-pair
are applied. The decay
rates for 
the unpolarized case of
the final $Z$ boson
are presented in the second column.
The last column results are for
the decay rates corresponding 
to the 
longitudinal polarization of
the
final $Z$ boson. Furthermore, 
in this Table, we show for the 
tree level 
(and full one-loop) decay widths
in the first (second) 
line, respectively. When we consider all 
generation of neutrinos, one should 
add to data by overall factor $3$.
The one-loop corrections are 
about $\sim 10\%$
contributions to the tree-level 
decay rates. We note that 
one-loop corrections are evaluated
as follows: 
\begin{eqnarray}
 \delta [\%] = \dfrac{\Gamma^{\text{Full}} 
 -\Gamma^{\text{Tree}} }{\Gamma^{\text{Tree}}}
 \times 100\%. 
\end{eqnarray}

%%%%%%%%%%%%%%%%%%%%%%%%%%%%%%%%%%%
\begin{table}[H]
\begin{center}
\begin{tabular}{l@{\hspace{2cm}}
c@{\hspace{2cm}}c}  
\hline \hline 
$m_{\nu_e\bar{\nu}_e}^{\textrm{cut}}$ [GeV]
& $\Gamma_{H\rightarrow Z \nu_e\bar{\nu}_e}$  [KeV]
& $\Gamma_{H\rightarrow Z_{L} \nu_e\bar{\nu}_e}$  [KeV]
\\ \hline \hline
%%%%%%%%%%%%%%%%%%%%%%%%%%%%%%%%%%%%%%%%%%%%%%%%%%%%%
$0$  &   $5.8177 $  & $2.2872 $ \\
     &   $6.4174 $  & $2.5061 $ 
\\ 
\hline
$5$  &   $5.7014 $  & $2.1736 $ \\
     &   $6.2902 $  & $2.3818 $ 
\\ 
\hline
$10$  &  $5.3401 $  & $1.8515 $ \\
     &   $5.8943 $  & $2.0293 $ 
\\ 
\hline
$20$  &  $3.7362 $  & $0.8389$ \\
     &   $4.1305 $  & $0.9201 $ 
\\ 
\hline\hline
\end{tabular}
\caption{\label{DRON} The decay rates
for on-shell Higgs decay into
$Z \nu_e\bar{\nu}_e$. The first (second) 
line, we show for the tree level 
(and full one-loop) decay widths, 
respectively.}
\end{center}
\end{table}

\begin{figure}[ht]
\centering
$\begin{array}{cc}
\hspace{-5.8cm}
\dfrac{d \Gamma_{H \rightarrow 
Z \nu_l\bar{\nu}_l}}
{d m_{\nu_l\bar{\nu}_l} } 
&
\hspace{-6.6cm}
\delta[\%]
\\
\includegraphics[width=8.5cm,height=8cm]
{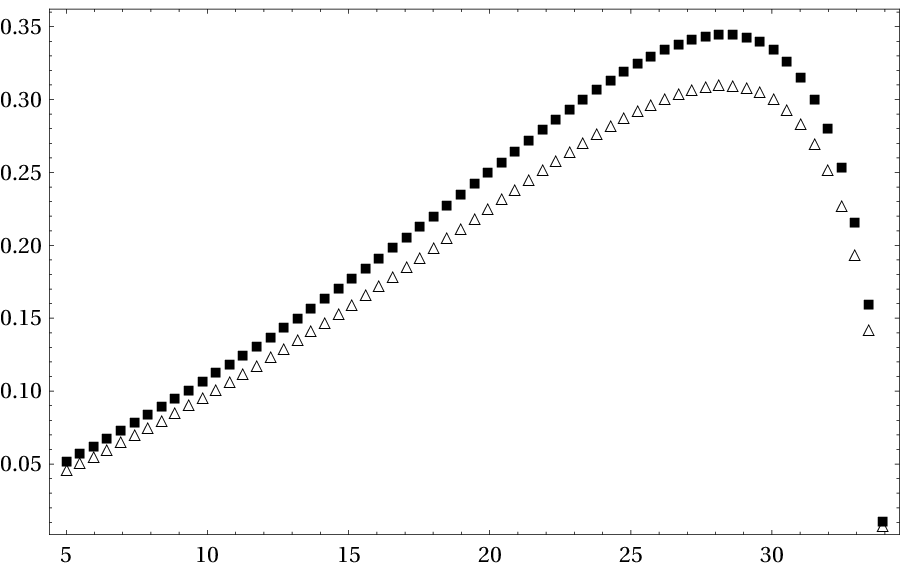}
& 
\includegraphics[width=8cm,height=8.1cm]
{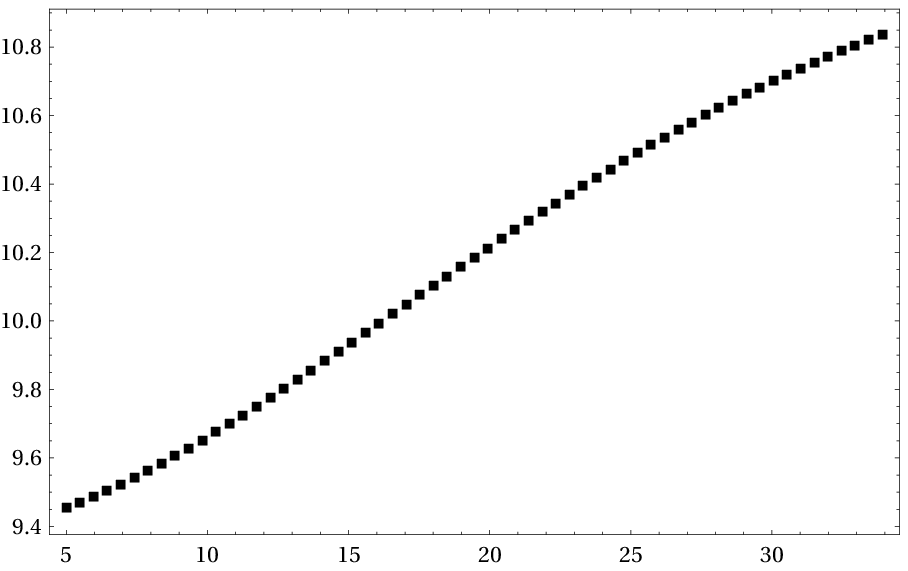}
\\
\hspace{6.5cm} m_{\nu_l\bar{\nu}_l} 
[\text{GeV}]
&
\hspace{6.2cm} m_{\nu_l\bar{\nu}_l} 
[\text{GeV}]
\end{array}$
\caption{\label{decay500} 
The differential decay rates 
(left panel) and corrections
(right panel) with 
respect to $m_{\nu_l\bar{\nu}_l}$
for the unpolarized $Z$ boson case.
In the left panel, 
the triangle points
show for the tree-level
decay widths and the 
rectangle points present for
full one-loop decay widths.
In the right panel, the electroweak
corrections are shown as the 
rectangle points.
}
\end{figure}

\begin{figure}[ht]
\centering
$\begin{array}{cc}
\hspace{-5.8cm}
\dfrac{d \Gamma_{H \rightarrow 
Z_L \nu_l\bar{\nu}_l}}
{d m_{\nu_l\bar{\nu}_l} } &
\hspace{-6.6cm}
\delta[\%]
\\
\includegraphics[width=8.5cm,height=8cm]
{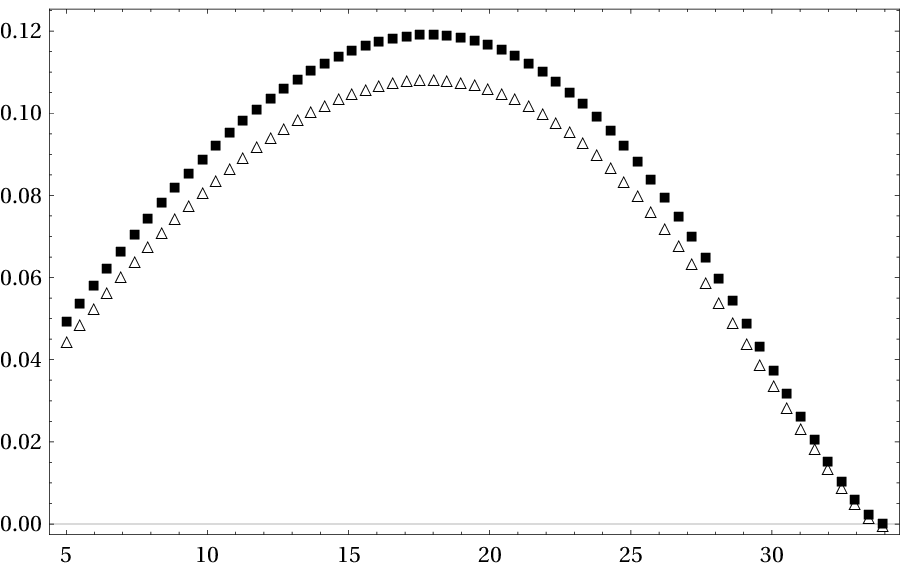}
& 
\includegraphics[width=8cm,height=8cm]
{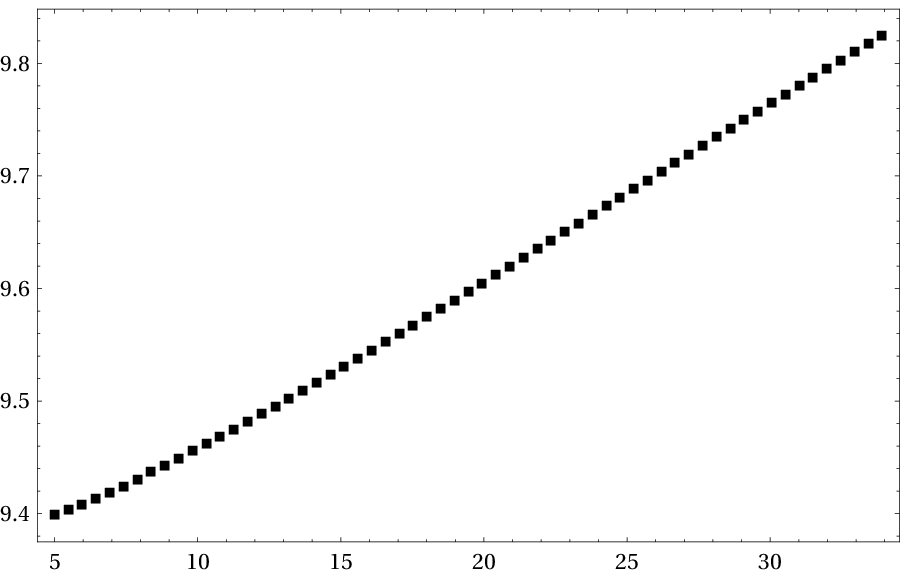}
\\
\hspace{6.5cm} m_{\nu_l\bar{\nu}_l} 
[\text{GeV}]
&
\hspace{6.2cm} m_{\nu_l\bar{\nu}_l} 
[\text{GeV}]
\end{array}$
\caption{\label{decay500LL}
The differential 
decay rates (left panel) and 
corrections (right panel) with 
respect to $m_{\nu_l\bar{\nu}_l}$
in the longitudinal polarization case
for $Z$ boson. 
In the left Figure, 
the triangle points present
for the tree-level
decay widths and the 
rectangle points are for
full one-loop decay widths.
In the right Figure, 
the electroweak corrections are 
plotted as the 
rectangle points.
}
\end{figure}

We next consider the off-shell
Higgs decay to $Z \nu_e\bar{\nu}_e$.
The numerical results are shown in 
the Table~\ref{DROFF}. In this case,
we only consider the unpolarized 
of $Z$ boson in the final state. 
In the first column, off-shell Higgs 
mass $M_{H^*}$ is shown 
in the range of 
$200$ GeV to $500$ GeV. The off-shell
decay widths are presented 
in the second
column in which the first (second)
line is for the tree-level 
(full one-loop) decay rates, 
respectively. It is worth to 
mention
that the results in off-shell
Higgs decays
are good agreement
with the decay rates 
in~\cite{Phan:2022amy}. 
This means that the main 
contributions
to the decay rates are 
from the values  
around the peak of $Z$-pole decay 
to $\nu_l\bar{\nu}_l$ 
(this explanation will be 
confirmed later).
%%%%%%%%%%%%%%%%%%%%%%%%%%%%%%%%%%%
\begin{table}[H]
\begin{center}
\begin{tabular}{c@{\hspace{2cm}}c}  
\hline \hline 
$M_{H^*}$ [GeV]
& $\Gamma_{H\rightarrow 
Z \nu_e\bar{\nu}_e}$  [GeV]
\\ \hline \hline
%%%%%%%%%%%%%%%%%%%%%%%%%%%%%%%%%%
$200$  &   $0.0478 $  \\
       &   $0.0541$  
     \\
\hline
$300$  &   $0.3383$ \\
       &   $0.3789$  
\\ 
\hline
$400$  &   $1.0124$  \\
     &     $1.1418$  
     \\
\hline
$500$  &   $2.2101$  \\
       &   $2.4865$  
     \\
\hline
\hline
\end{tabular}
\caption{\label{DROFF} 
The decay rates
for off-shell Higgs decay into
$Z \nu_e\bar{\nu}_e$.
The first (second) 
line, we show for the tree level 
(and full one-loop) decay widths, 
respectively.
}
\end{center}
\end{table}
%%%%%%%%%%%%%%%%%%%%%%%%%%%%%%%%
For the experimental analyses,
differential decay rates with 
respect to the invariant mass of 
neutrino-pair are of interests. 
These are corresponding
to the decay rates of
Higgs decay to $Z$ plus missing 
energy. Thus, the data will 
provide  
the precise backgrounds for 
the signals of Higgs decay to 
lepton-pair when $Z\rightarrow$
lepton-pair is taken into account. 
This also contributes to the signals
of $H\rightarrow$
invisible particles if the decay of 
final $Z$ boson to neutrino-pair
is considered. 
In Fig.~\ref{decay500},
we show for the differential 
decay rates with 
respect to $m_{\nu_l\bar{\nu}_l}$
for the case of the 
unpolarized $Z$ final state.  
We apply a cut of $m_{\nu_l
\bar{\nu}_l}^{\text{cut}}
\ge 5$ GeV for this study. 
In the left panel, the triangle points
are for the tree-level
decay widths and the 
rectangle points are
of
full one-loop decay widths.
In the right panel, the electroweak
corrections are plotted. One finds
that the corrections are range 
of $9.4\%$ to $10.8\%$ contributions. 
%%%%%%%%%%%%%%%%%%%%%%%%%%%%%%%%%%%%%%%%%
In Fig.~\ref{decay500LL},
the same distributions are shown 
in the longitudinal polarization 
of final $Z$ boson. We use the same 
convention as previous case. 
We also find the corrections 
are range of 
$9.4\%$ to $9.8\%$ contributions. 
%%%%%%%%%%%%%%%%%%%%%%%%%%%%%%%%
%%%%%%%%%%%%%%%%%%%%%%%%%%%%%%%%%%%%%
%%%%%%%%%%%%%%%%%%%%%%%%%%%%%%%%%%

\begin{figure}[H]
\centering
$\begin{array}{cc}
\hspace{-5.8cm}
\dfrac{d \Gamma_{H \rightarrow 
Z \nu_l\bar{\nu}_l}}
{d m_{\nu_l\bar{\nu}_l} } 
&
\hspace{-6.6cm}
\delta[\%]
\\
\includegraphics[width=8.5cm,height=8cm]
{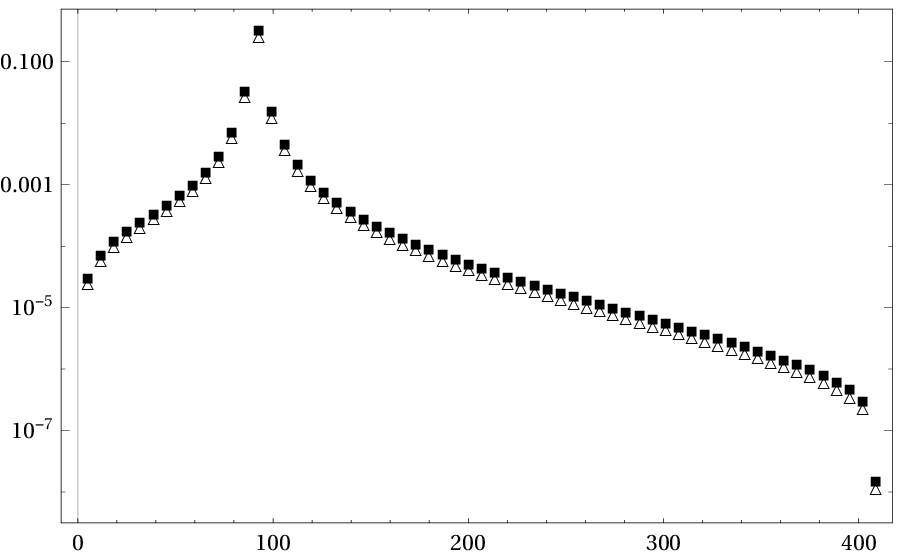}
& 
\includegraphics[width=8cm,height=8.1cm]
{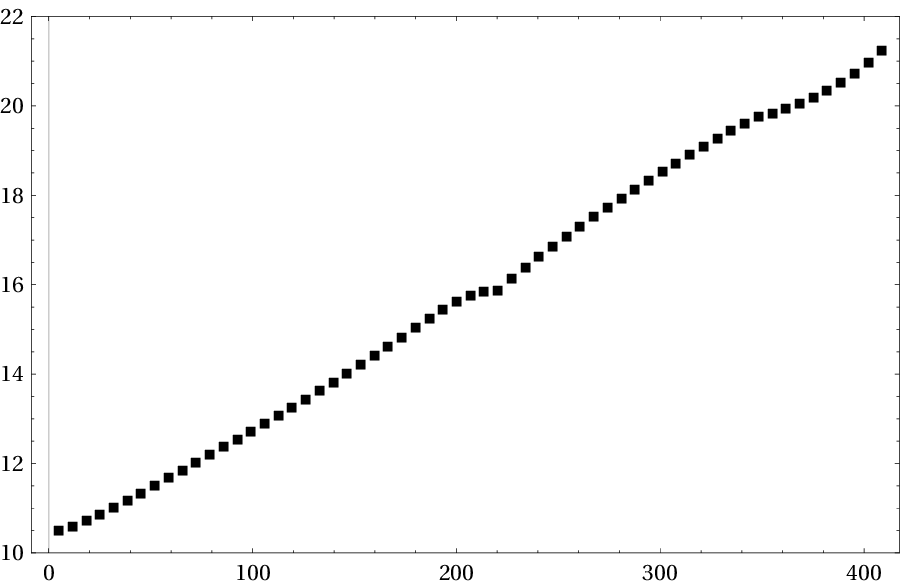}
\\
\hspace{6.5cm} m_{\nu_l\bar{\nu}_l} 
[\text{GeV}]
&
\hspace{6.2cm} m_{\nu_l\bar{\nu}_l} 
[\text{GeV}]
\end{array}$
\caption{\label{decay500OFF} 
The differential 
decay rates (left panel) and 
corrections (right panel)
with 
respect to $m_{\nu_l\bar{\nu}_l}$ 
for off-shell
Higgs case.
In the left panel, 
tree-level
decay widths are plotted as 
triangle points 
and full one-loop decay widths
are shown as rectangle points.
In the right panel, 
the electroweak corrections are 
presented as the 
rectangle points.
}
\end{figure}

\begin{figure}[ht]
\hspace{1.0cm}
\centering
\includegraphics[width=8.0cm,
height=5.0cm]
{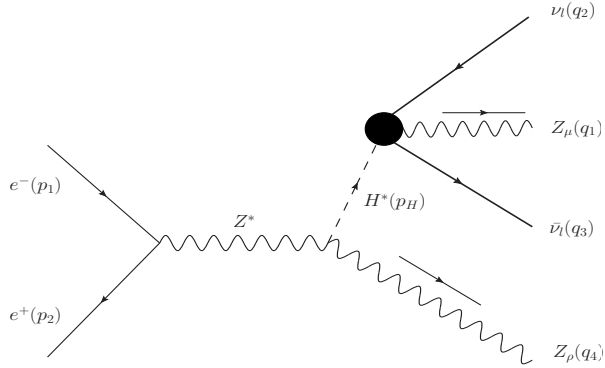}
\centering
\caption{\label{signalDiagram} 
Feynman diagram for the processes
$e^-e^+ \rightarrow Z(Z\nu_l\bar{\nu_l})$ 
at the ILC with the blob representing for 
one-loop corrections to $H\rightarrow
Z\nu_l\bar{\nu_l}
$.} 
\end{figure}

The differential 
decay rates with 
respect to $m_{\nu_l\bar{\nu}_l}$ 
for off-shell
Higgs case at $M_{H}^* =500$ GeV are
generated. In the 
Figs.~\ref{decay500OFF}, we observe 
a peak at $m_{\nu_l\bar{\nu}_l} = M_Z$
which is corresponding to $Z\rightarrow 
\nu_l\bar{\nu}_l$. 
The decay rates give large values 
around the peak and
fall down rapidly beyond the peak. 
The corrections are from $10\%$
to $25\%$ in all range of
$m_{\nu_l\bar{\nu}_l}$. We note that 
a cut of $m_{\nu_l\bar{\nu}_l}^{\text{cut}}
\ge 5$ GeV is employed in the distribution. 
From the distribution, it is shown that
the main contributions to 
the off-shell Higgs 
decay rates come from the corresponding
values around $Z$-peak. 
It explains that the off-shell Higgs 
decay rates in this work are good
agreement with the results in~\cite{Phan:2022amy}. 
This convinces the previous 
conclusion about the data in 
Table~\ref{DROFF}. 
For all range of Higgs mass, 
we also check numerically
that the dominant contributions to the 
decay rates come from the $Z$-pole diagrams,
or the diagrams of $H\rightarrow Z Z^*
\rightarrow Z \nu_l\bar{\nu}_l$ (from group
$1$ and $2$) in these decay
channels. The same conclusion has pointed
out in the paper~\cite{Kachanovich:2021pvx}.
%%%%%%%%%%%%%%%%%%%%%%%%%%%%%%%
%%%%%%%%%%%%%%%%%%%%%%%%%%%%%%%%
We turn our attention to 
analyse the signals 
$H\rightarrow 
Z \nu_l\bar{\nu}_l$ through 
Higgs productiuon at future 
lepton collider such as 
$e^-e^+ \rightarrow 
ZH^* \rightarrow Z 
(Z \nu_l\bar{\nu}_l)$
with including the
initial beam 
polarizations. Differential 
cross section
with resprect to $M_{H^*}$ 
is given by~\cite{Phan:2022amy}:
\begin{eqnarray}
\label{masterformulas1}
\dfrac{d\sigma^{e^-e^+\rightarrow
ZH^*\rightarrow Z(Z \nu_l\bar{\nu}_l)}
(\sqrt{s}) }{dM_{H^*}}
&=&(2M_{H^*}^2)\times 
\dfrac{ 
\sigma^{e^-e^+\rightarrow ZH^*}(\sqrt{s}, M_{H^*})}
{[(M_{H^*}^2-M_H^2)^2+ \Gamma_H^2
M_H^2 ]}\times \dfrac{\Gamma_{H^*\rightarrow ZZ}
(M_{H^*})}{\pi}.
\end{eqnarray}
Feynman diagram is depicted in 
the Fig.~(\ref{signalDiagram}). 
%%%%%%%%%%%%%%%%%%%%%%%%%%%%%%%
We mention that 
cross section for $e^-e^+ \rightarrow ZH^*$ 
can be found in \cite{Phan:2022amy}.
Total cross section for 
these processes can be computed as
follows:
\begin{eqnarray}
\sigma^{e^-e^+\rightarrow
ZH^*\rightarrow Z(Z \nu_l\bar{\nu}_l)}
= \int\limits_{M_Z}^{\sqrt{s}-M_Z}
d M_{H^*} \; 
\dfrac{d\sigma^{e^-e^+\rightarrow
ZH^*\rightarrow Z(Z \nu_l\bar{\nu}_l)}
(\sqrt{s}) }{dM_{H^*}}.
\end{eqnarray}
In Table~\ref{sigeeHZvlvl}, we show
cross sections for the signals 
of Higgs decay to $Z \nu_l\bar{\nu}_l$ 
via 
$e^-e^+ \rightarrow 
ZH^* \rightarrow Z 
(Z \nu_l\bar{\nu}_l)$
with including the
initial beam 
polarizations (taking all
three generations of neutrinos in the data).
The second (third) 
column
presents for the signals at tree
level (full correction) cross sections 
respectively. The last column
is for the SM backgrounds 
which are tree level 
of the reactions $e^-e^+ \rightarrow 
Z Z \nu_l\bar{\nu}_l$. The background 
processes are generated 
by using GRACE~\cite{Belanger:2003sd}. 
At each 
center-of-mass energy, the first line
shows for LR case and second line
is for RL polarization case. 
We show that the signals
$H\rightarrow  Z\nu_l\bar{\nu}_l$
can be probed at center-of-mass 
energy $\sqrt{s}=250$ GeV and 
these are hard to 
measure at higher-energy 
regions due to the dominant of
the backgrounds.
%%%%%%%%%%%%%%%%%%%%%%%%%%%%%%%%%%%%%%
\begin{table}[h]
\begin{center}
\begin{tabular}{l@{\hspace{2cm}}l@{\hspace{2cm}}l
@{\hspace{2cm}}l}  
\hline \hline 
$\sqrt{s}$ [GeV] & $\sigma_{\text{sig}}^{\text{Tree}}$ [fb] 
                 & $\sigma_{\text{sig}}^{\text{Full}}$ [fb]  
                 & $\sigma_{\text{bkg}}$ [fb]
\\ \hline \hline
$250$  & $2.43873$    & $2.69398$   & $0.00309$ \\ 
       & $1.58487$    & $1.74649$   & $0.00016$ \\    \hline  
$500$  & $0.68498$    & $0.75668$   & $16.7839$  \\ 
       & $0.44404$    & $0.48932$   & $1.33409$  \\   \hline  
$1000$ & $0.26879$    & $0.29692$   & $164.146$  \\ 
       & $0.17424$    & $0.19201$   & $1.16635$  \\   \hline \hline        
%%%%%%%%%%%%%%%%%%%%%%%%%%%%%%%%%%%%%%%%%%%%%%%%
\end{tabular}
\caption{\label{sigeeHZvlvl} Total cross section
of $e^-e^+ \rightarrow Z(Z\nu_l\bar{\nu_l})$. 
First line results show for $LR$ of $e^-e^+$, 
second line results present for $RL$ 
of $e^-e^+$. Tree generations for neutrinos
are taken into the results.}
\end{center}
\end{table}
%%%%%%%%%%%%%%%%%%%%%%%%%%%%%%%%%%%

\begin{figure}[H]
\centering
$\begin{array}{cc}
\hspace{-4.5cm}
\dfrac{d \sigma_{LR}
}
{d M_{H^*} } \Big[\dfrac{\textrm{fb}}{\textrm{GeV}}\Big]
&
\hspace{-4.5cm}
\dfrac{d \sigma_{RL}
}
{d M_{H^*} } 
\Big[\dfrac{\textrm{fb}}{\textrm{GeV}}\Big]
\\
\includegraphics[width=8cm,height=8cm]
{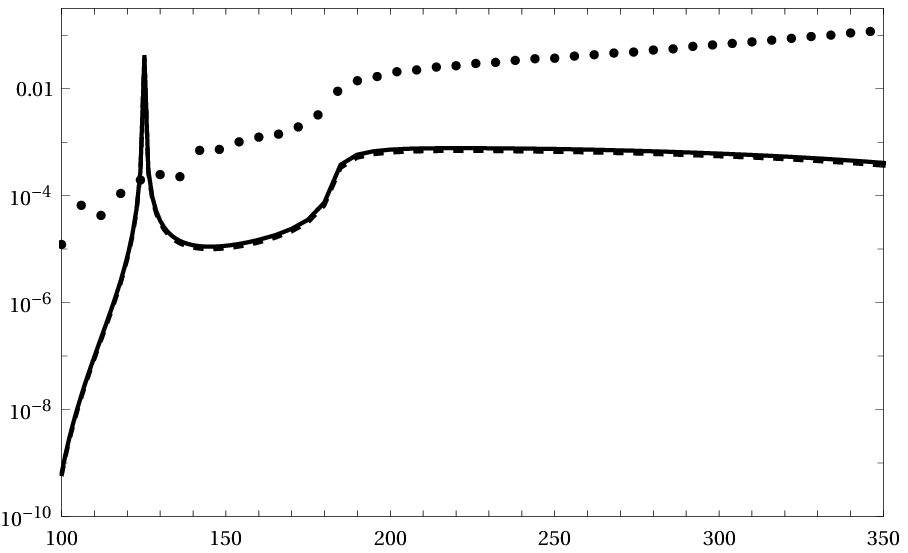}
& 
\includegraphics[width=8cm,height=8cm]
{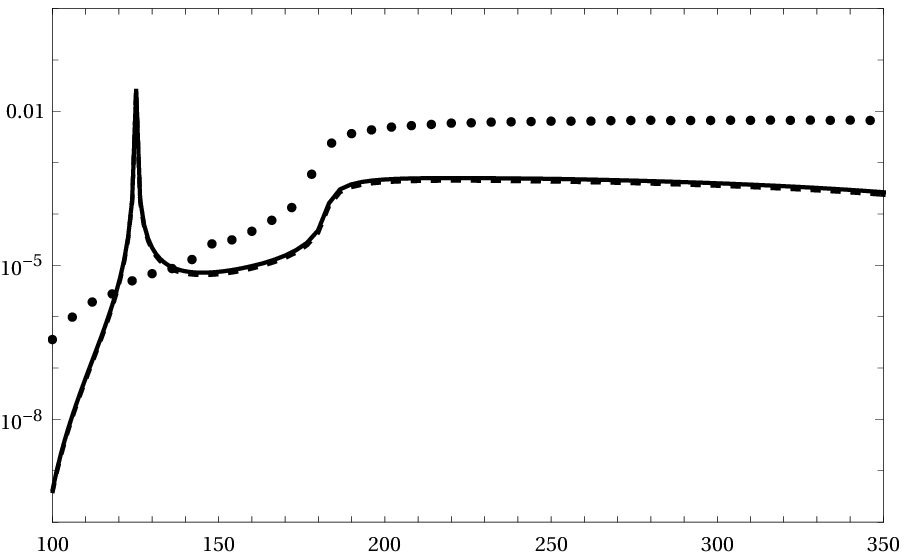}
\\
\hspace{6.2cm} M_{H^*}
[\text{GeV}]
&
\hspace{6.2cm} M_{H^*} 
[\text{GeV}]
\end{array}$
\caption{\label{cross500all} 
Off-shell Higgs 
decay rates as a function of 
$M_{H^*}$ at center-of-mass 
energy $\sqrt{s}=500$ GeV.
Three generations for neutrinos
are included in the results.
Cross sections
for LR case are shown in the left panel 
and for RL are presented in the 
right panel. For the signal cross sections,
tree-level cross sections are shown as 
dashed line
and full one-loop cross sections
are shown as solid line.
The dotted points are for 
the SM backgrounds.
}
\end{figure}

In the Fig.~\ref{cross500all}, we 
plot the distributions for cross 
section as functions of $M_{H^*}$ 
at $\sqrt{s}=500$ GeV of 
center-of-mass energy, considering
the 
initial polarization cases for $e^-e^+$. 
Cross sections
for LR case are shown in the left panel 
and for RL are presented in the 
right panel. 
For the signal cross sections,
tree-level cross sections are plotted 
as dashed line
and full one-loop cross sections
are presented as solid line. While
the SM backgrounds are shown
as dotted points.
The off-shell Higgs mass $M_{H^*}$ 
is varied from $M_Z$ 
to $\sqrt{s}-M_Z$. 
It is observed that the cross section
are dominant around the on-shell Higgs
mass $M_{H^*}\sim 125$ GeV. 
It is well-known that 
we have another peak which is around the 
ZH threshold ($\sim M_Z+M_H= 215$ GeV). 
Due to the small
value of total decay width of Higgs boson,
on-shell Higgs mass peak becomes more 
visible than the later one.
In the off-shell 
Higgs mass region, cross sections are 
much smaller 
(about $2$ order smaller) than the ones 
around the on-shell Higgs mass peak.
We observe that the signals are 
clearly visible at on-shell Higgs mass 
$M_{H^*}=125$ GeV.
In the 
off-shell Higgs mass region, 
the SM backgrounds 
are much larger than the signals.
These large contributions are
mainly  
attributed to the dominant of 
$t$-channel diagrams appear 
in the background processes.
%%%%%%%%%%%%%%%%%%%%%%%%%%%%%%%%%%%

Full one-loop electroweak corrections
to the process $e^-e^+ \rightarrow ZH$
and the SM background processes
with 
including the initial beam polarizations
should be taken into account for the above 
analyses. The corrections can be generated
by using the program~\cite{Belanger:2003sd}, 
and recently
study in \cite{Quach:2022rjy}. Furthermore, 
by generalizing the couplings of Nambu-Goldstone
bosons to Higgs, gauge bosons, etc as 
in \cite{Phan:2021pcc}, 
we can extend our work for many beyond the SM. 
These topics will be addressed in our future works. 
%%%%%%%%%%%%%%%%%%%%%%%%%%%%%
\section{Conclusions}   %%%%%
%%%%%%%%%%%%%%%%%%%%%%%%%%%%% 
Analytical results for one-loop
contributing to the decay 
processes
$H\rightarrow 
Z \nu_l\bar{\nu}_l$ for 
$l=e,\mu, \tau$ 
in 't Hooft-Veltman gauge 
have presented. The calculations 
have performed within the Standard 
Model framework. One-loop form factors 
are expressed in terms of 
the Passarino-Veltman functions
in the standard conventions 
of {\tt LoopTools} which the decay rates
can be evaluated numerically. 
We have also studied 
the signals of $H\rightarrow 
Z \nu_l\bar{\nu}_l$ through Higgs 
productions at future lepton collider
such as $e^-e^+ \rightarrow 
ZH^* \rightarrow Z 
(Z \nu_l\bar{\nu}_l)$
with including the initial beam 
polarizations. The SM background 
processes for this analysis
have also taken into account. 
In phenomenological results, 
we find that one-loop corrections
are about $10\%$ contributions 
to the decay rates. They are sizeable
contributions and should be included
at future colliders. 
We show that the signals 
$H\rightarrow  Z\nu_l\bar{\nu}_l$
are clearly visible at center-of-mass 
energy $\sqrt{s}=250$ GeV and 
these are 
hard to probe at higher-energy 
regions due to the dominant of
the background.
\\

\noindent
{\bf Acknowledgment:}~
This research is funded by 
Vietnam National Foundation 
for Science and Technology 
Development (NAFOSTED) under 
the grant number $103.01$-$2019.346$.\\ 
%
%%%%%%%%%%%%%%%%%%%%%%%%%%%%%%%%
\section*{Appendix $A$: 
Tensor reduction}%%         
%%%%%%%%%%%%%%%%%%%%%%%%%%%%%%%%
We show all tensor one-loop 
reduction formulas which
have applied for this 
calculation in this appendix.
The technique is based 
on the method in~\cite{Denner:2005nn}. 
Tensor one-loop one-, 
two-, three- and four-point 
integrals with rank $R$ 
are defined:
\begin{eqnarray}
\label{tensoroneloop}
\{A; B; C; D\}^{\mu_1\mu_2\cdots \mu_R} 
= (\mu^2)^{2-d/2}
\int \frac{d^dk}{(2\pi)^d} 
\dfrac{k^{\mu_1}k^{\mu_2}
\cdots k^{\mu_R}}{\{P_1; P_1 P_2;P_1P_2P_3; 
P_1P_2P_3P_4\}}.
\end{eqnarray}
Where the 
inverse Feynman propagators
$P_j$ ($j=1,\cdots, 4$) 
are given by
\begin{eqnarray}
 P_j = (k+ q_j)^2 -m_j^2 +i\rho.
\end{eqnarray}
In this definition, the momenta 
$q_j = \sum\limits_{i=1}^j p_i$ with 
$p_i$ for the external momenta are 
taken into account and   
$m_j$ for internal masses in the loops.
The internal masses can be real and complex
in the calculation. Following  
the dimensional regularization method, 
one-loop integrals are peformed 
in space-time dimension 
$d=4-2\varepsilon$. 
The renormalization
scale is introduced as
$\mu^2$ in this definition 
that help to track of the correct 
dimension of the integrals in 
space-time dimension 
$d$.
If the numerators of one-loop integrands
in Eq.~(\ref{tensoroneloop}) are $1$,
we have the corresponding scalar
one-loop functions (noted 
as $A_0$, $B_0$, $C_0$ and $D_0$). 
All reduction formulas for 
one-loop tensor integrals 
up to rank $R=3$ are are presented
in the following paragraphs. 
In detail, one has the reduction 
expressions for one-loop two-point 
tensor integrals as
%%%%%%%%%%%%%%%%%%%%%%%%%%%%%%%%%%%
\begin{eqnarray}
A^{\mu}        &=& 0, \\
A^{\mu\nu}     &=& g^{\mu\nu} A_{00}, \\
A^{\mu\nu\rho} &=& 0,\\
B^{\mu}        &=& q^{\mu} B_1,\\
B^{\mu\nu}     &=& g^{\mu\nu} B_{00} + q^{\mu}q^{\nu} B_{11}, \\
B^{\mu\nu\rho} &=& \{g, q\}^{\mu\nu\rho} B_{001} 
+ q^{\mu}q^{\nu}q^{\rho} B_{111}, 
\end{eqnarray}
%%%%%%%%%%%%%%%%%%%%%%%%%%%%%%%%%%%
Reduction formulas for one-loop
tensor three-point integrals are 
shown as
\begin{eqnarray}
C^{\mu}        &=& q_1^{\mu} C_1 + q_2^{\mu} C_2 
 = \sum\limits_{i=1}^2q_i^{\mu} C_i, 
\\
C^{\mu\nu}    &=& g^{\mu\nu} C_{00} 
 + \sum\limits_{i,j=1}^2q_i^{\mu}q_j^{\nu} C_{ij},
\\
C^{\mu\nu\rho} &=&
\sum_{i=1}^2 \{g,q_i\}^{\mu\nu\rho} 
C_{00i}+
\sum_{i,j,k=1}^2 q^{\mu}_i q^{\nu}_j q^{\rho}_k C_{ijk},
\end{eqnarray}
%%%%%%%%%%%%%%%%%%%%%%%%%%%%%%%%%%%
For four-point functions, 
we have simillarly
reduction expressions: 
\begin{eqnarray}
D^{\mu}        &=& q_1^{\mu} D_1 + q_2^{\mu} D_2 + q_3^{\mu}D_3 
 = \sum\limits_{i=1}^3q_i^{\mu} D_i, \\
 D^{\mu\nu}    &=& g^{\mu\nu} D_{00} 
 + \sum\limits_{i,j=1}^3q_i^{\mu}q_j^{\nu} D_{ij}, 
\\
D^{\mu\nu\rho} &=&
	\sum_{i=1}^3 \{g,q_i\}^{\mu\nu\rho} D_{00i}+
	\sum_{i,j,k=1}^3 q^{\mu}_i q^{\nu}_j q^{\rho}_k D_{ijk}.
\end{eqnarray}
%%%%%%%%%%%%%%%%%%%%%%%%%%%%%%%%%%%
We have already used the short notation~\cite{Denner:2005nn} 
$\{g, q_i\}^{\mu\nu\rho}$ which is 
written explicitly 
as follows: $\{g, q_i\}^{\mu\nu\rho} = g^{\mu\nu} q^{\rho}_i 
+ g^{\nu\rho} q^{\mu}_i + g^{\mu\rho} q^{\nu}_i$. It is noted that all 
scalar coefficients  $A_{00}, B_1, \cdots, D_{333}$
in the right hand sides of the 
above reduction formulas 
are so-called Passarino-Veltman functions
~\cite{Denner:2005nn}. These functions have 
implemented into {\tt LoopTools}~\cite{Hahn:1998yk} for 
numerical computations. 
%%%%%%%%%%%%%%%%%%%%%%%%%%%%%%%%%%%%%%%%%%
\section*{Appendix $B$: Numerical checks}  
%%%%%%%%%%%%%%%%%%%%%%%%%%%%%%%%%%%%%%%%%%%
After having all the neccessary one-loop 
form factors, we are going to check 
the computation numerically. We find that
$F_{00}$ contains the $UV$-divergent. 
By taking 
the one-loop counter term which are 
corresponding to the $F_{00}^{(G_4)}$. 
The analytic expressions for 
$F_{00}^{(G_4)}$ are given 
in (\ref{CTF4}) in which all 
renormalization 
constants are shown 
in the appendix $D$.

In the Table~(\ref{UV}), 
checking for 
the UV-finiteness of the 
results at a random point in 
phase space are presented. By varying 
$C_{UV}$ parameters the amplitudes are good 
stability over more than $14$ digits.
%%%%%%%%%%%%%%%%%%%%%%%%%%%%%%%%%%%%%%
\begin{table}[ht]
\begin{center}
\begin{tabular}{l@{\hspace{2cm}}l}  
\hline \hline 
$(C_{UV}, \mu^2)$
& 2 $\mathcal{R}$e$
\{M_{\text{Tree}}^*M_{\text{1-Loop}}\}$  
\\ \hline \hline
%%%%%%%%%%%%%%%%%%%%%%%%%%%%%%%%%%%%%%%%%%%%
$(0, 1)$ &  $-0.0015130298318390845 
- 0.001513160592122863\; i $\\ \hline 
%%%%%%%%%%%%%%%%%%%%%%%%%%%%%%%%%%%%%%%%%%%%
$(10^2, 10^5)$ &$ -0.0015130298318393881 
- 0.001513160592122863\; i$\\ \hline 
%%%%%%%%%%%%%%%%%%%%%%%%%%%%%%%%%%%%%%%%%%%%
 $(10^4, 10^{10})$ & $-0.0015130298318233315 
 - 0.001513160592122863\; i$ \\
\hline\hline
\end{tabular}
\caption{\label{UV} Checking for 
the UV-finiteness of the 
results at an random point in phase space.
The amplitude $M_{\text{1-Loop}}$ is included
all one-loop diagrams and counterterm diagrams.
}
\end{center}
\end{table}
%%%%%%%%%%%%%%%%%%%%%%%%%%%%%%%%%%%%%%%%%%%
\section*{Appendix $C$: Self energy}     %%         
%%%%%%%%%%%%%%%%%%%%%%%%%%%%%%%%%%%%%%%%%%%
All Self energy are presented 
in terms of PV- functions in 
't Hooft-Veltman gauge. 
%%%%%%%%%%%%%%%%%%%%%%%%%%%%%%%%%%%
\subsection*{Self energy $A$-$A$}%%         
%%%%%%%%%%%%%%%%%%%%%%%%%%%%%%%%%%%
Self-energy photon-photon 
functions
are casted into two 
fermion and contributions
as follows:
\begin{eqnarray}
\Pi^{AA}(q^2)=
\Pi^{AA}_{T,b} (q^2)+\Pi^{AA}_{T,f} (q^2). 
\end{eqnarray}
Each part is given:
\begin{eqnarray}
\Pi^{AA}_{T,b} (q^2)
&=&
\dfrac{e^2}{(4\pi)^2}
\Bigg\{
\big(4 M_W^2+3 q^2\big) 
B_0(q^2,M_W^2,M_W^2)
-2 (d-2) A_0(M_W^2)
\Bigg\}
, \\
%%%%%%%%%%%%%%%%%%%%%%%%%%%%%%%%%%%
\Pi^{AA}_{T,f} (q^2)
&=&
\dfrac{e^2}{(4 \pi)^2}
\Bigg\{
- 2 \sum_f N^C_f Q_f^2 
\Big[4 B_{00}(q^2,m_f^2,m_f^2)
% \n \\
% &&\hspace{5.0cm}
+ q^2 B_0(q^2,m_f^2,m_f^2)
- 2 A_0(m_f^2)\Big]
\Bigg\}
.\n\\
\end{eqnarray}
%%%%%%%%%%%%%%%%%%%%%%%%%%%%%%%%%%%
\subsection*{Self energy $Z$-$A$}%%         
%%%%%%%%%%%%%%%%%%%%%%%%%%%%%%%%%%%
Self-energy functions 
for $Z$-$A$ mixing are 
written as the same previous 
form. Each part is presented
accordingly
\begin{eqnarray}
\Pi^{ZA}_{T,b} (q^2)
&=&
\frac{e^2}{(32 \pi^2) (d-1) s_W c_W}
\Bigg\{
2 (d-2) \Big[c_W^2 (2 d-3)-s_W^2\Big] 
A_0(M_W^2)
\\
&&
- \Big\{
4 M_W^2 \Big[c_W^2 (3 d-4)+(d-2) s_W^2\Big]
+q^2 \Big[c_W^2 (6 d-5)+s_W^2\Big]
\Big\}
B_0(q^2,M_W^2,M_W^2)
\Bigg\}
, \n \\
%%%%%%%%%%%%%%%%%%%%%%%%%%%%%%%%%%%
\Pi^{ZA}_{T,f} (q^2)
&=&
\dfrac{e^2}{(32 \pi ^2) s_W c_W}
\Bigg\{
2 \sum_f N^C_f Q_f 
\Big(2 s_W^2 Q_f-T^3_f\Big) 
\times
\\
&&\hspace{4.5cm} \times
\Big[4 B_{00}(q^2,m_f^2,m_f^2)
+q^2 B_0(q^2,m_f^2,m_f^2)
-2 A_0(m_f^2)\Big]
\Bigg\}
. \n
\end{eqnarray}
%%%%%%%%%%%%%%%%%%%%%%%%%%%%%%%%%%%
\subsection*{Self energy $Z$-$Z$}%%         
%%%%%%%%%%%%%%%%%%%%%%%%%%%%%%%%%%%
Self energy functions for 
$Z$-$Z$ are shown in terms of 
scalar one-loop integrals as follows:
\begin{eqnarray}
\Pi^{ZZ}_{T,b} (q^2)
&=&
\dfrac{e^2}{(64 \pi^2) (d-1) q^2 s_W^2 c_W^4}
\Bigg\{
2 q^2 c_W^2 (2-d) 
\Big[c_W^4 (4 d-7)+ s_W^2 (s_W^2-2 c_W^2)\Big] A_0(M_W^2)
\n \\
&&
+c_W^2 \Big[M_H^2-M_Z^2-(d-2) q^2\Big] A_0(M_H^2)
+c_W^2 \Big[M_Z^2-M_H^2-(d-2) q^2\Big] A_0(M_Z^2)
\n \\
&&
+ \Big\{
2 q^2 
\Big[c_W^2 (M_H^2 + M_Z^2)-2 M_W^2 (d-1)\Big]
-c_W^2 \Big[ (M_H^2-M_Z^2)^2 + q^4 \Big]
\Big\}
B_0(q^2,M_H^2,M_Z^2) 
\n \\
&&
+
\Big\{4 M_W^2 
\Big[
(3 c_W^4 - s_W^4) (2 d-3)
-2 c_W^2 s_W^2
\Big]
\\
&&
+q^2 \Big[
3 c_W^4 (4 d-3)
+(2 c_W^2 - s_W^2) s_W^2
\Big]
\Big\} c_W^2 q^2 B_0(q^2,M_W^2,M_W^2)
\Bigg\}
, \n \\
%%%%%%%%%%%%%%%%%%%%%%%%%%%%%%%%%%%
\Pi^{ZZ}_{T,f} (q^2)
&=&
\dfrac{e^2}{(16 
\pi ^2) s_W^2 c_W^2}
\sum_f N^C_f 
\times
\\
&&\times
\Bigg\{
\Big[(T^3_f)^2 \big(2 m_f^2-q^2\big)
+2 q^2 Q_f s_W^2 
\big( T^3_f - Q_f s_W^2 \big)
\Big] 
B_0(q^2,m_f^2,m_f^2)
\n \\
&&
+ \Big[4 Q_f s_W^2 (T^3_f - Q_f s_W^2)
- 2 (T^3_f)^2\Big] \Big[ 2B_{00}(q^2,m_f^2,m_f^2)
- A_0(m_f^2) \Big]
\Bigg\}
. \n 
\end{eqnarray}
%%%%%%%%%%%%%%%%%%%%%%%%%%%%%%%%%%%
\subsection*{Self energy $W$-$W$}%%         
%%%%%%%%%%%%%%%%%%%%%%%%%%%%%%%%%%%
Self-energy functions for 
$W$-$W$ are presented correspondingly
\begin{eqnarray}
\Pi^{WW}_{T,b} (q^2)
&=&
\dfrac{e^2}{(64 \pi^2) (d-1) q^2 s_W^2 c_W^2}
\Bigg\{
c_W^2 \Big[M_H^2-M_W^2-(d-2) q^2\Big] A_0(M_H^2)
\n \\
&&
+c_W^2 
\Big[
2 M_W^2
-M_H^2
-M_Z^2
-2 q^2 (2 d-3) (d-2)
\Big] A_0(M_W^2)
\n \\
&&
+c_W^2 \Big[4 c_W^2 (d-2)+1\Big] 
\Big[M_Z^2-M_W^2-(d-2) q^2\Big] A_0(M_Z^2)
\\
&&
+ \Big\{
c_W^2 q^4 \Big[4 c_W^2 (3 d-2)-1\Big]
-c_W^2 (M_W^2-M_Z^2)^2 \Big[4 c_W^2 (d-2)+1\Big]
\n \\
&&
+ 2 q^2 M_W^2 
\Big[2 c_W^4 (3 d-5)
-2 s_W^4 (d-1)
+3 c_W^2 (2 d-3)+1
\Big]
\Big\} B_0(q^2,M_W^2,M_Z^2)
\n \\
&&
+c_W^2 
\Big\{
2 q^2 \Big[(3-2 d) M_W^2+M_H^2\Big]
-(M_H^2-M_W^2)^2
-q^4
\Big\} B_0(q^2,M_H^2,M_W^2)
\n \\
&&
+4 c_W^2 s_W^2 
\Big\{
M_W^2 (2 q^2 - M_W^2) (d-2)+(3 d-2) q^4
\Big\} B_0(q^2,0,M_W^2)
\Bigg\}
, \n \\
%%%%%%%%%%%%%%%%%%%%%%%%%%%%%%%%%%%
\Pi^{WW}_{T,f} (q^2)
&=&
\dfrac{e^2}{(64 \pi^2) s_W^2 c_W^2}
\Bigg\{
2 c_W^2
\sum_{\text{doublet}} N^C_f 
\Big[
\Big(m_f^2 + m_{f'}^2 
- q^2 \Big) B_0(q^2,m_{f'}^2,m_f^2)
\\
&&\hspace{5cm}
-4 B_{00}(q^2,m_{f'}^2,m_f^2)
+A_0(m_f^2)
+A_0(m_{f'}^2)
\Big]
\Bigg\}
.\n
\end{eqnarray}
%%%%%%%%%%%%%%%%%%%%%%%%%%%%%%%%%%%
\subsection*{Self energy $H$-$H$}%%         
%%%%%%%%%%%%%%%%%%%%%%%%%%%%%%%%%%%
The expressions for 
self-energy $H$-$H$
are written
\begin{eqnarray}
\Pi^{HH}_{b} (q^2)
&=&
\dfrac{e^2}{(128 \pi^2) 
M_W^2 s_W^2 c_W^4}
\Bigg\{
3 M_H^2 c_W^4 
\Big[
3 M_H^2 B_0(q^2,M_H^2,M_H^2)
+ A_0(M_H^2) \Big]
\n \\
&&
+ 2 c_W^4 
\Big\{
4 M_W^2 \Big[M_W^2 (d-1) - q^2\Big]
+M_H^4
\Big\} B_0(q^2,M_W^2,M_W^2)
\\
&&
+ \Big\{
c_W^4 M_H^4+4 M_W^2 \Big[M_W^2 (d-1) - c_W^2 q^2\Big]
\Big\} B_0(q^2,M_Z^2,M_Z^2)
\n \\
&&
+2 c_W^4 \Big[2 M_W^2 (d-1) 
+ M_H^2\Big] A_0(M_W^2)
+ \Big[c_W^4 M_H^2 + 
2 M_W^2 c_W^2 (d-1)\Big] 
A_0(M_Z^2)
\Bigg\}
, \n \\
%%%%%%%%%%%%%%%%%%%%%%%%%%%%%%%%%%%
\Pi^{HH}_{f} (q^2)
&=&
\dfrac{e^2}{(128 \pi ^2) 
M_W^2 s_W^2 c_W^4}
\times
\n \\
&&\times
\Bigg\{
4 c_W^4 \sum_f N^C_f m_f^2 
\Big[(q^2-4 m_f^2) B_0(q^2,m_f^2,m_f^2)
-2 A_0(m_f^2)\Big]
\Bigg\}
- \dfrac{3 \delta T}{v}
.
\end{eqnarray}
Where $v=246$ GeV is 
vacuum expectation value.
%%%%%%%%%%%%%%%%%%%%%%%%%%%%%%%%%%%
\subsection*{The tadpole}        %%         
%%%%%%%%%%%%%%%%%%%%%%%%%%%%%%%%%%%
The tadpole is calculated as follows:
\begin{eqnarray}
T^{loop}_b
&=&
\dfrac{e}{(64 \pi^2) M_W s_W c_W^2}
\Bigg\{
\Big[c_W^2 M_H^2+2 M_W^2 (d - 1)\Big]
A_0(M_Z^2) 
\\
&&\hspace{3.0cm}
+2 c_W^2 \Big[2 M_W^2 (d-1) 
+ M_H^2\Big] A_0(M_W^2)
+3 M_H^2 c_W^2 A_0(M_H^2)
\Bigg\}
, \n \\
%%%%%%%%%%%%%%%%%%%%%%%%%%%%%%%%%%%
T^{loop}_f
&=&-
\dfrac{ 8 e\; c_W^2}{(64 \pi ^2) M_W s_W c_W^2}
\sum_f N^C_f m_f^2 A_0(m_f^2)
.
\end{eqnarray}
We then have
\begin{eqnarray}
\delta T = - (T^{loop}_b + T^{loop}_f).
\end{eqnarray}
In case of neutrino, 
explicit expressions for
self-energy functions 
$\nu_l$-$\nu_l$ 
as follows
\begin{eqnarray}
\Sigma^{\nu_l} (q^2)
&=&
\mathcal{K}^{\nu_l}_{\gamma} (q^2) \slashed{q}
+
\mathcal{K}^{\nu_l}_{5\gamma} (q^2) \slashed{q} \gamma_5
\end{eqnarray}
where 
\begin{eqnarray}
\mathcal{K}^{\nu_l}_{\gamma} (q^2)
=
- \mathcal{K}^{\nu_l}_{5\gamma} (q^2)
&=&
-\dfrac{e^2}{128\pi^2 s_W^2 c_W^2}
\Bigg[
(2 c_W^2+1)
+
2 B_{1}(q^2,0,M_Z^2)
\\
&&\hspace{3.5cm}
+2 c_W^2
\sum_l
\Big(\frac{m_l^2}{M_W^2}+2\Big) 
B_{1}(q^2,m_l^2,M_W^2)
\Bigg].
\nonumber
\end{eqnarray}
%%%%%%%%%%%%%%%%%%%%%%%%%%%%%%%%%%%%%%%
\section*{Appendix $D$: Counterterms}%%         
%%%%%%%%%%%%%%%%%%%%%%%%%%%%%%%%%%%%%%%
The counterterms of the decay process 
$H \rightarrow Z \nu_l \bar{\nu_l}$ 
are written by 
\begin{eqnarray}
\label{CTF4}
%%%%%%%%%%%%%%%%%%%%%%%%%%%%%%%%%%%
F_{00}^{(G_4)} 
&=& 
F_{00,Z \nu_l \bar{\nu_l}}^{(G_4)}
+
F_{00,HZZ}^{(G_4)}
+
F_{00,Z \chi_3}^{(G_4)}
+
F_{00,Z Z}^{(G_4)}
,
\end{eqnarray}
where
\begin{eqnarray}
F_{00,Z \nu_l \bar{\nu_l}}^{(G_4)}
&=&
\dfrac{2 \pi \alpha M_W}{s_W^2 c_W^3}
\dfrac{1}{s-M_Z^2 + i \Gamma_Z M_Z}
\Big(
\delta Y
+
\delta G_2
+
\delta G_3
+
\delta Z_{ZZ}^{1/2}
+
2 \delta Z_{\nu_l L}^{1/2}
\Big)
, \\
F_{00,HZZ}^{(G_4)}
&=&
\dfrac{2 \pi \alpha M_W}{s_W^2 c_W^3}
\dfrac{1}{s-M_Z^2 + i \Gamma_Z M_Z}
\Big(
\delta Y
+
\delta G_2
+
\delta G_3
+
\delta G_Z
+
2 \delta Z_{ZZ}^{1/2}
+
\delta Z_H^{1/2}
\Big)
, \\
F_{00,Z Z}^{(G_4)}
&=&
\frac{2 \pi \alpha M_W}{s_W^2 c_W^3}
\dfrac{1}{(s-M_Z^2 + i \Gamma_Z M_Z)^2}
\Big(
2 M_Z^2 \, \delta G_Z
+
(M_Z^2-s) \delta Z_{ZZ}^{1/2}
\Big). 
\end{eqnarray}
While the contribution of 
$F_{00,Z \chi_3}^{(G_4)}$ is 
vanished due to Dirac equation.

All renormalization constants are given 
as 
\begin{eqnarray}
&& \delta Y = 
-\delta Z _{AA} ^{1/2} 
+ \dfrac{s_W}{c_W} \delta Z_{ZA} ^{1/2},
\\
&& \delta G_2
= \delta G_Z - \delta H 
, \quad 
\delta G_3
= \delta G_Z - \delta G_W,
\\
&& \delta H
=
\dfrac{\delta M_Z^2 - \delta M_W^2}{2(M_Z^2 - M_W^2)}
, \quad 
\delta G_Z
= 
\dfrac{\delta M_Z^2}{2 M_Z^2}
, \quad 
\delta G_W
= 
\dfrac{\delta M_W^2}{2 M_W^2}.
\end{eqnarray}
Other renormalization constants 
are read as
\begin{eqnarray}
&&\delta Z _{AA} ^{1/2} = 
\dfrac{1}{2} \dfrac{d}{d q^2} \Pi^{AA}_T (0)
= \dfrac{1}{2} \dfrac{d}{d q^2} 
\Pi^{AA}_T (q^2) \Big|_{q^2 = 0},
\\
&&\delta Z_{ZA} ^{1/2} = 
- \Pi^{ZA}_T (0)/M_Z^2 = 
- \Pi^{ZA}_T (q^2)/M_Z^2 \Big|_{q^2 = 0},
\\
&& \delta M_W^2
=
- \mathcal{R}\text{e} \, 
\Big\{ \Pi ^{WW}_T (M_W^2)
\Big\}
=
- \mathcal{R}\text{e} 
\, \Big\{ \Pi ^{WW}_T (q^2)
\Big|_{q^2 = M_W^2} \Big\},
\\
&& \delta M_Z^2
=
- \mathcal{R}\text{e} \, 
\Big\{
\Pi ^{ZZ}_T (M_Z^2)
\Big\}
=
- \mathcal{R}\text{e} \, 
\Big\{ \Pi ^{ZZ}_T (q^2) 
\Big|_{q^2 = M_Z^2} \Big\},
\\
&& \delta Z_{ZZ}^{1/2}
=
\frac{1}{2}
\mathcal{R}\text{e} \, 
\Big\{
\dfrac{d}{d q^2} 
\Pi ^{ZZ}_T (q^2) \Big|_{q^2 = M_Z^2}
\Big\}
=
\frac{1}{2}
\mathcal{R}\text{e} \, 
\Big\{
\Pi ^{ZZ'}_T (q^2) \Big|_{q^2 = M_Z^2}
\Big\},
\\
&& \delta Z_H^{1/2}
=
- \frac{1}{2}
\mathcal{R}\text{e} \,
\Big\{
\dfrac{d}{d q^2}
\Pi^{HH} (q^2) \Big|_{q^2 = M_H^2} 
\Big\}
=
- \frac{1}{2}
\mathcal{R}\text{e} \,
\Big\{
\dfrac{d}{d q^2}
\Pi^{HH} (q^2) \Big|_{q^2 = M_H^2} 
\Big\},
%%%%%%%%%%%%%%%%%%%%%%%%%%%%%%%%
%&& \delta Z_{f L}^{1/2}
%=
%\frac{1}{2}
%\mathcal{R}\text{e} \,
%\Big\{
%\mathcal{K}^f_{5\gamma} (m_{f}^2)
%-
%\mathcal{K}^f_{\gamma} (m_{f}^2)
%\Big\}
%- m_f 
%\dfrac{d}{d q^2}
%\Big\{
%m_f
%\mathcal{R}\text{e} \,
%\mathcal{K}^f_{\gamma} (q^2)
%+
%\mathcal{R}\text{e} \,
%\mathcal{K}^f_{1} (q^2)
%\Big\}
%\Big|_{q^2 = m_f^2}
%%%%%%%%%%%%%%%%%%%%%%%%%%%%%%%%
\\
&& 
\delta Z_{\nu_l L}^{1/2}
=
\frac{1}{2}
\mathcal{R}\text{e} \,
\Big\{
\mathcal{K}^{\nu_l}_{5\gamma} (m_{\nu_l}^2)
-
\mathcal{K}^{\nu_l}_{\gamma} (m_{\nu_l}^2)
\Big\}
.
\end{eqnarray}
%%%%%%%%%%%%%%%%%%%%%%%%%%%%%%%%%%%%%%%%%%%
\section*{Appendix $E$: Feynman diagrams}%%         
%%%%%%%%%%%%%%%%%%%%%%%%%%%%%%%%%%%%%%%%%%%
All Feynman diagrams contributing to 
the decay processes $H\rightarrow 
Z \nu_l\bar{\nu}_l$ in 't Hooft-Veltman 
are shown in this appendix. 
%%%%%%%%%%%%
\begin{figure}[ht]
\centering
\includegraphics[width=8.0cm, height=5.0cm]
{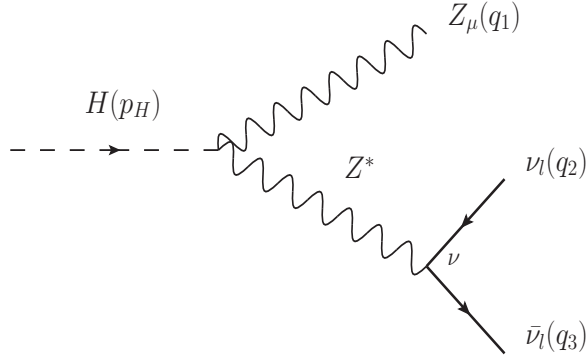}
\centering
\caption{Group $G_0$: Tree level 
Feynman diagram.}
\end{figure}
%%%%%%%%%%%%%%%%%%%%%%%%%%%%%%
\begin{figure}[ht]
\centering
\includegraphics[width=16.0cm, height=10.0cm]
{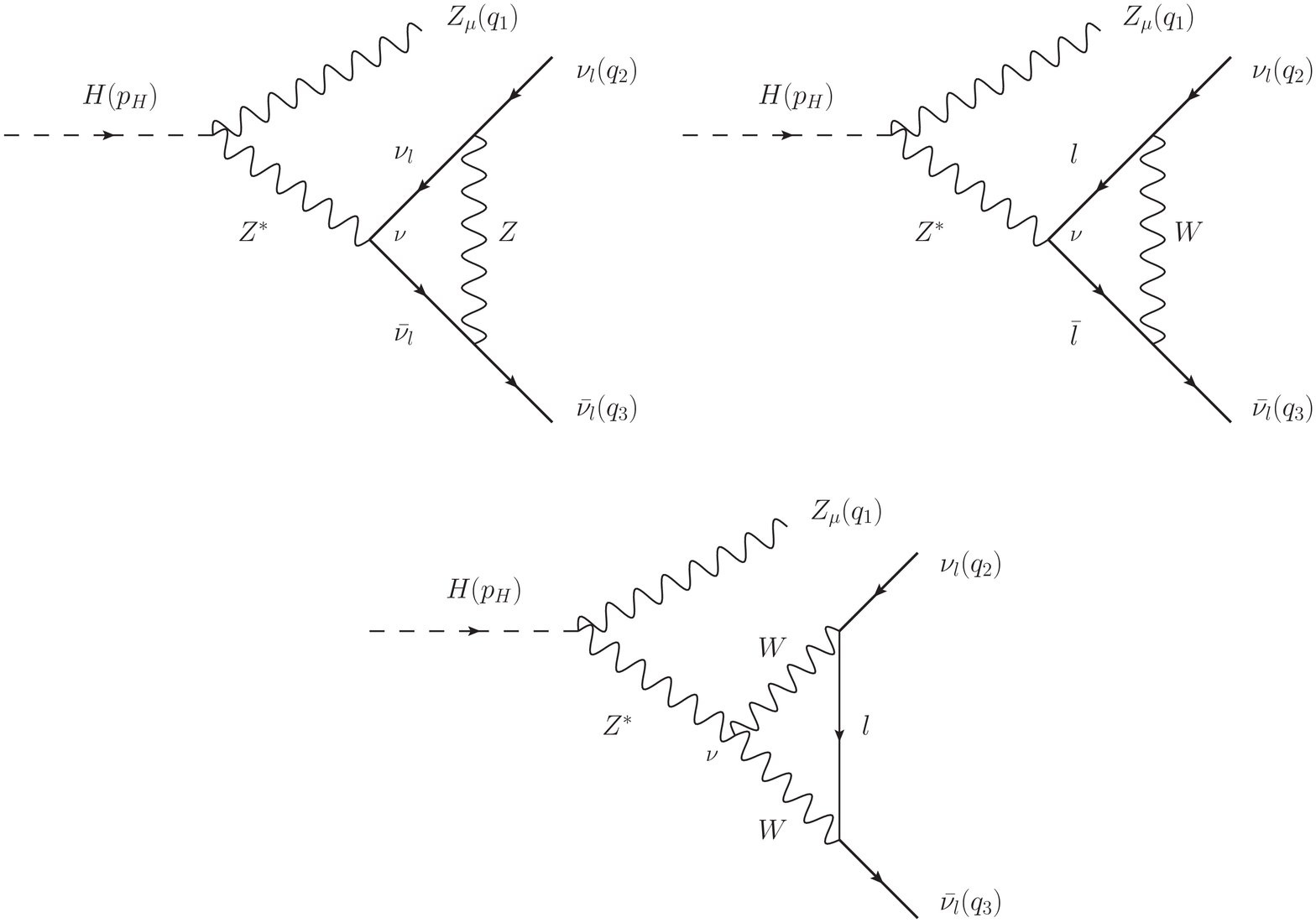}
\centering
\caption{Group $G_1$: All one-loop 
Feynman diagrams contributing to the vertex
$H\nu_l\bar{\nu}_l$.}
\end{figure}
%%%%%%%%%%%%%%%%%%%%%%%%%%%%%%
\begin{figure}[ht]
\centering
\includegraphics[width=15.0cm, height=3.2cm]
{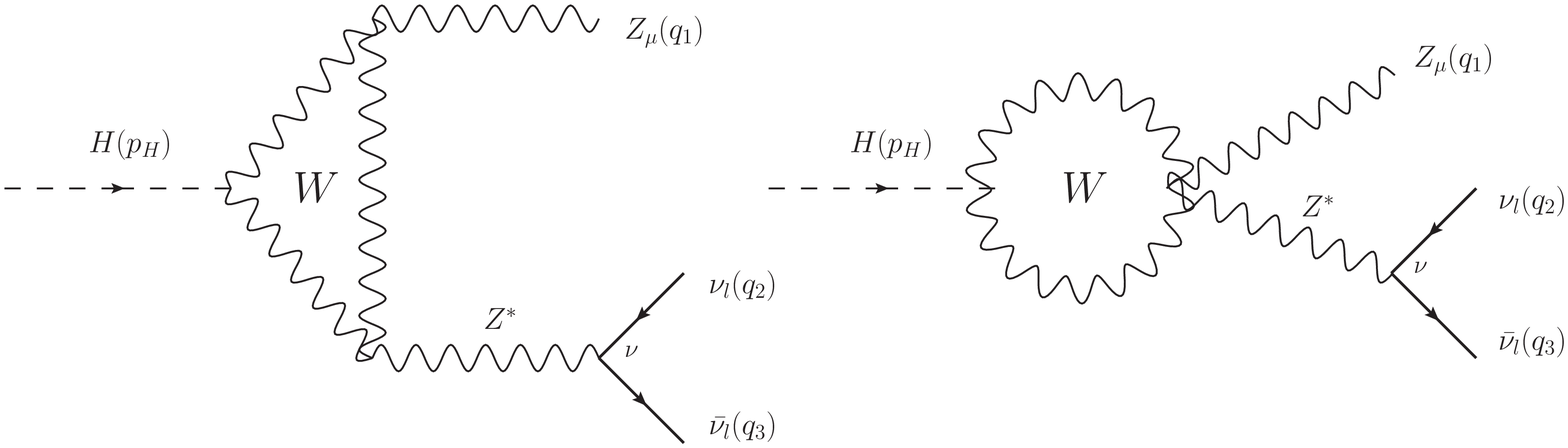}
\centering
\includegraphics[width=15.0cm, height=3.2cm]
{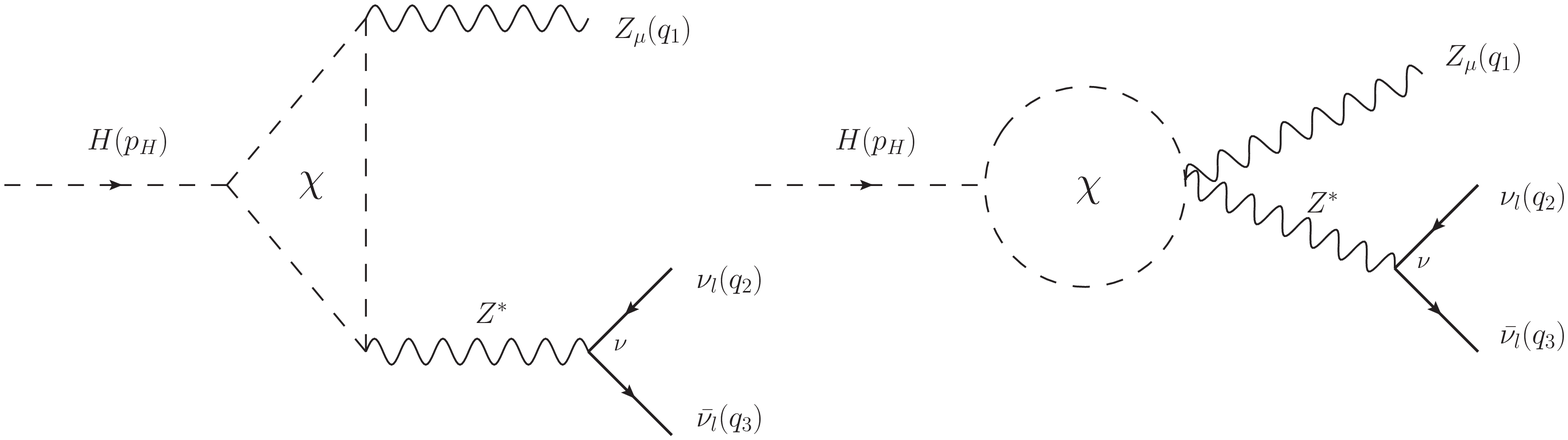}
\centering
\includegraphics[width=15.0cm, height=3.2cm]
{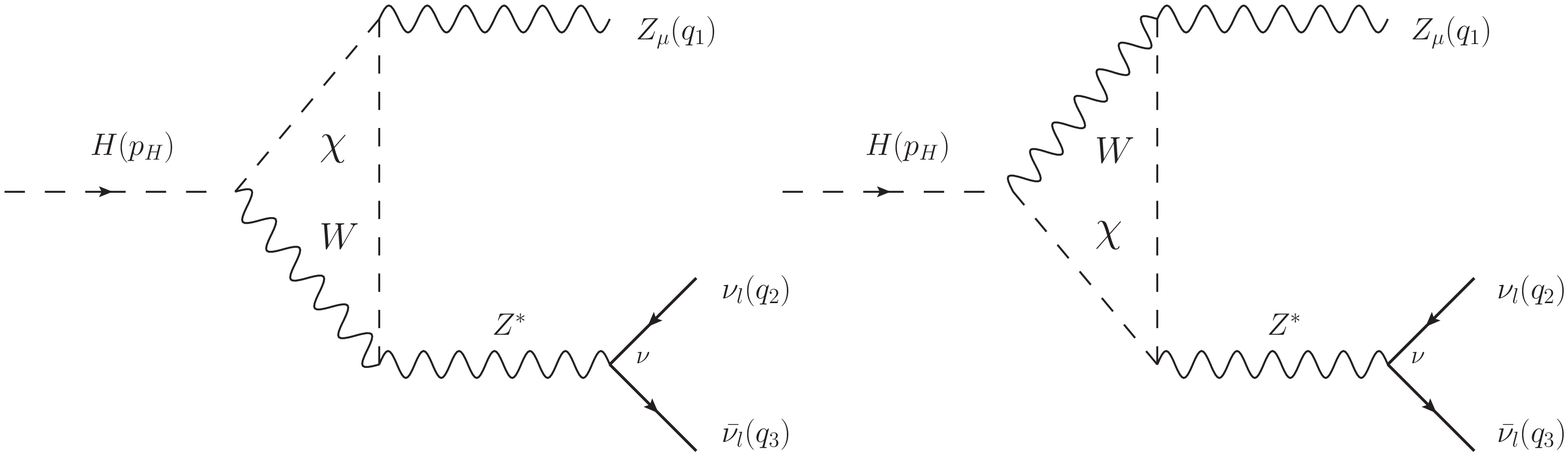}
\centering
\includegraphics[width=15.0cm, height=3.2cm]
{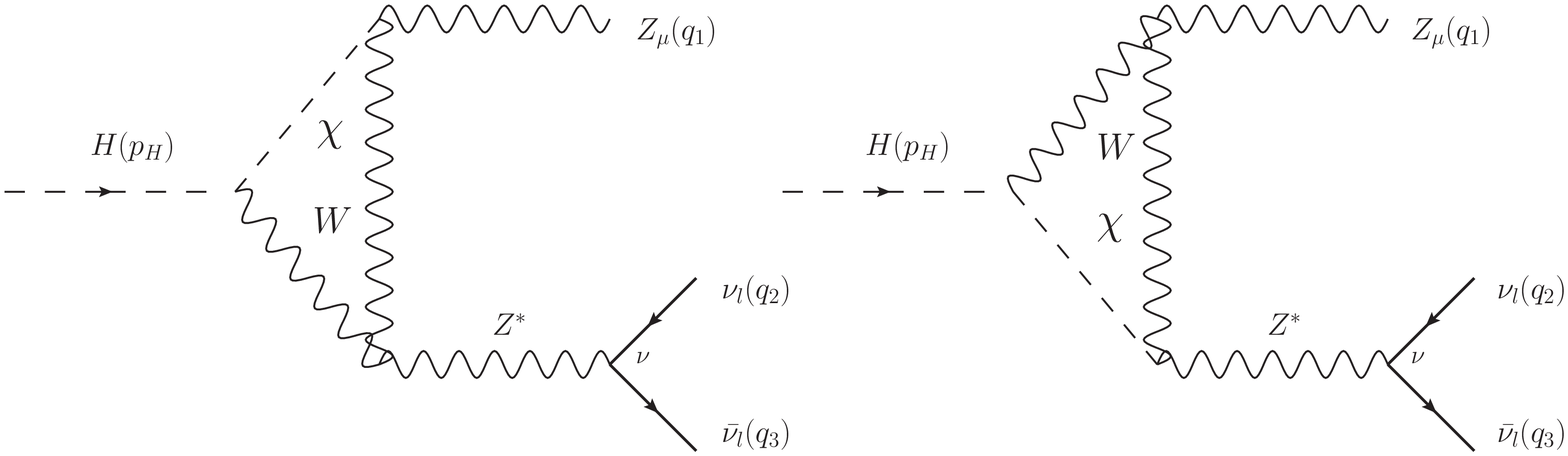}
\centering
\includegraphics[width=15.0cm, height=3.2cm]
{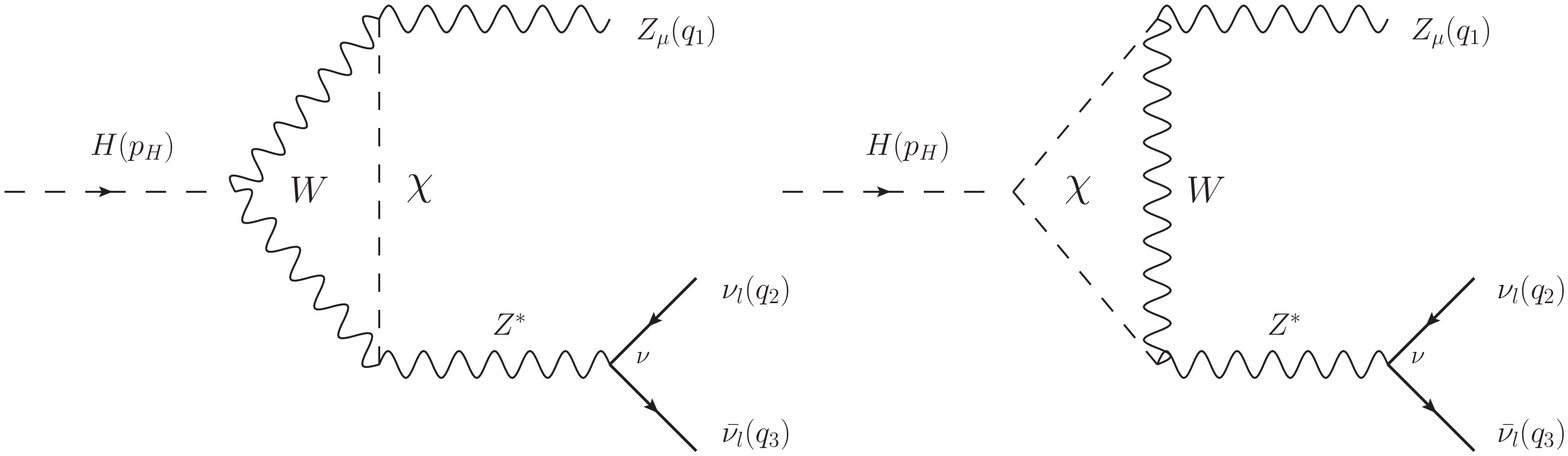}
\centering
\includegraphics[width=15.0cm, height=3.2cm]
{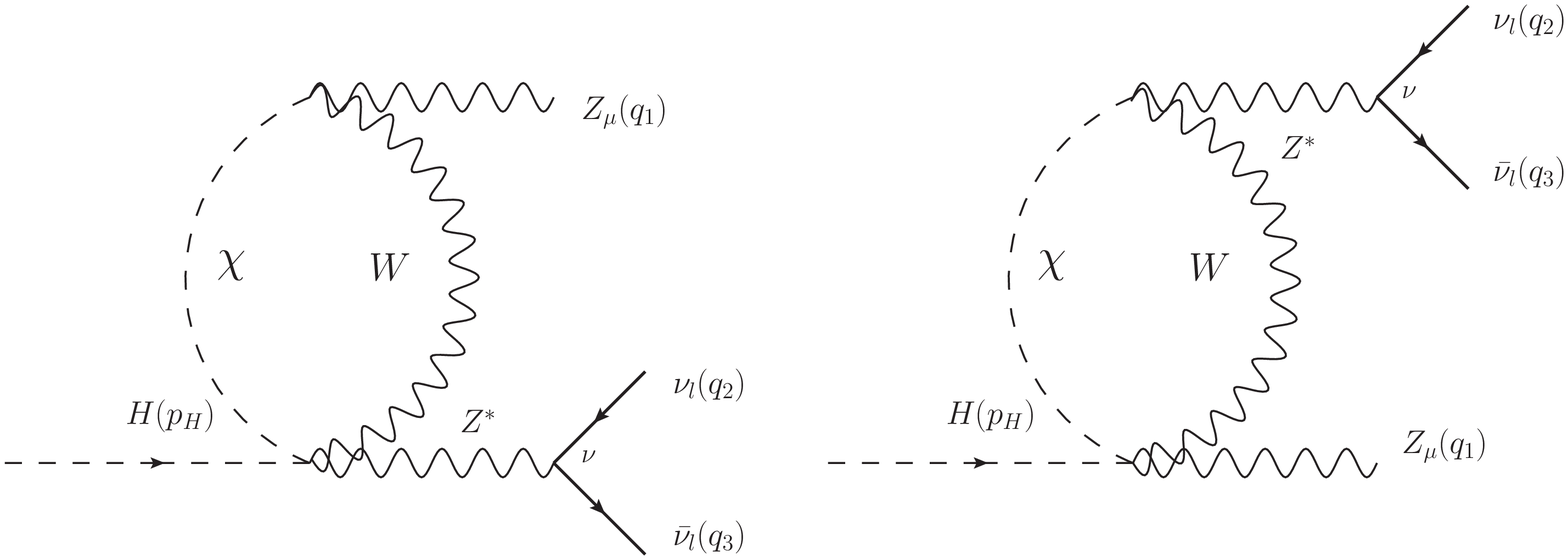}
\centering
\includegraphics[width=15.0cm, height=3.2cm]
{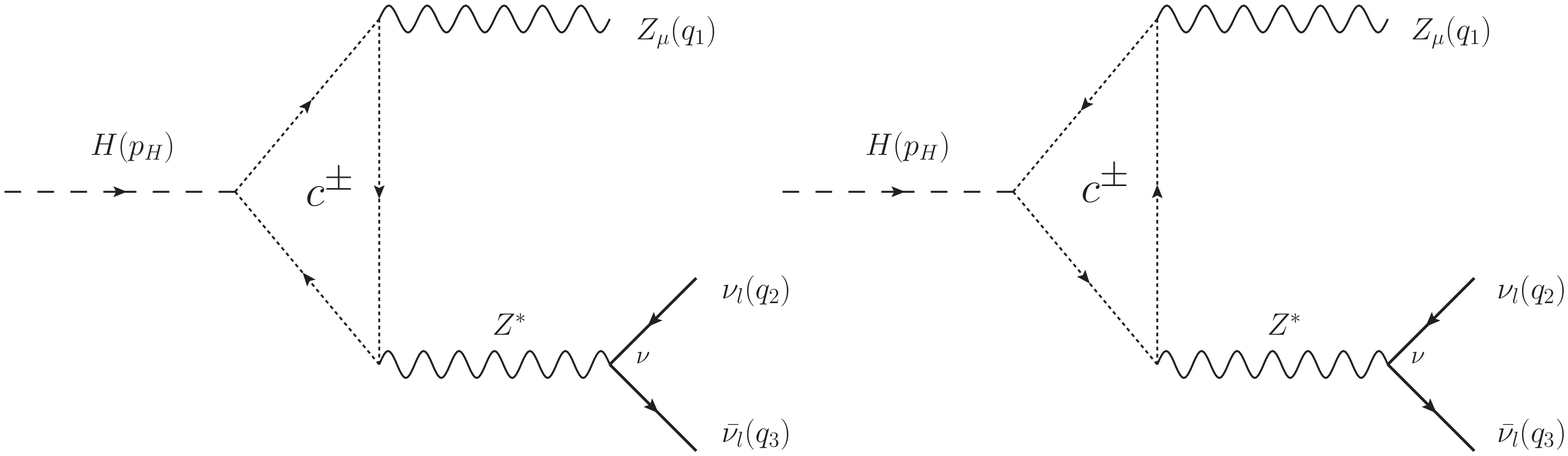}
\centering
\caption{Group $G_2$: All $Z$-pole Feynman
diagrams contributing to the decay process.
We note that $\chi^{\pm}$ and $c^{\pm}$
are Nambu-Goldstone bosons and 
ghost particles, respectively.}
\end{figure}
%%%%%%%%%%%%%%%%%%%%%%%%%%
\begin{figure}[ht]
\centering
\includegraphics[width=15.0cm, height=3.2cm]
{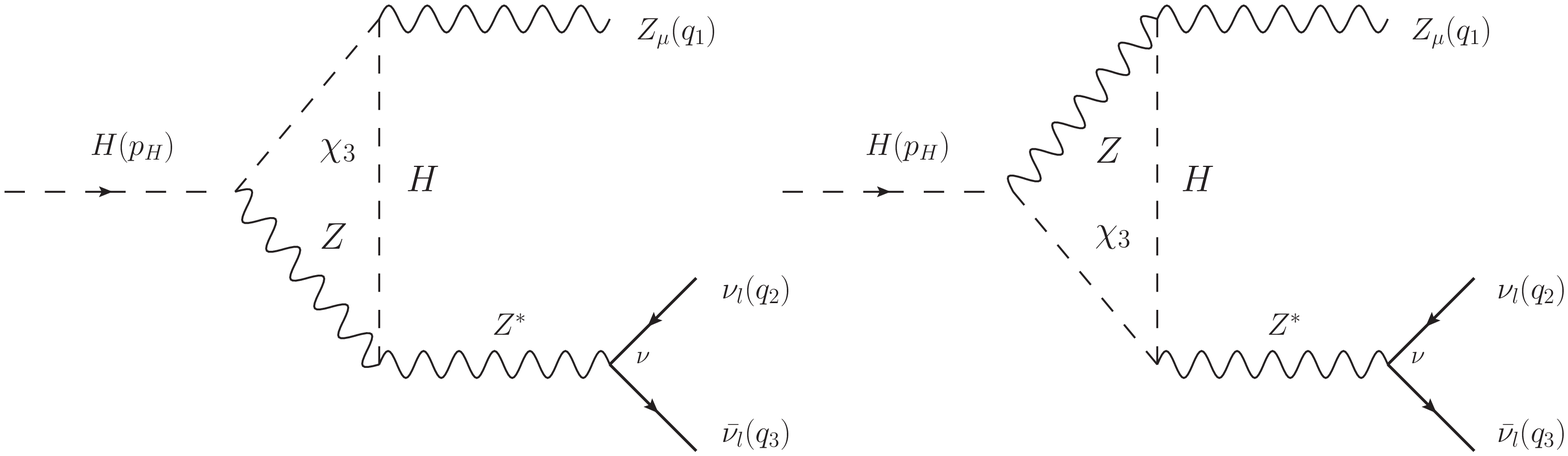}
\centering
\includegraphics[width=15.0cm, height=3.2cm]
{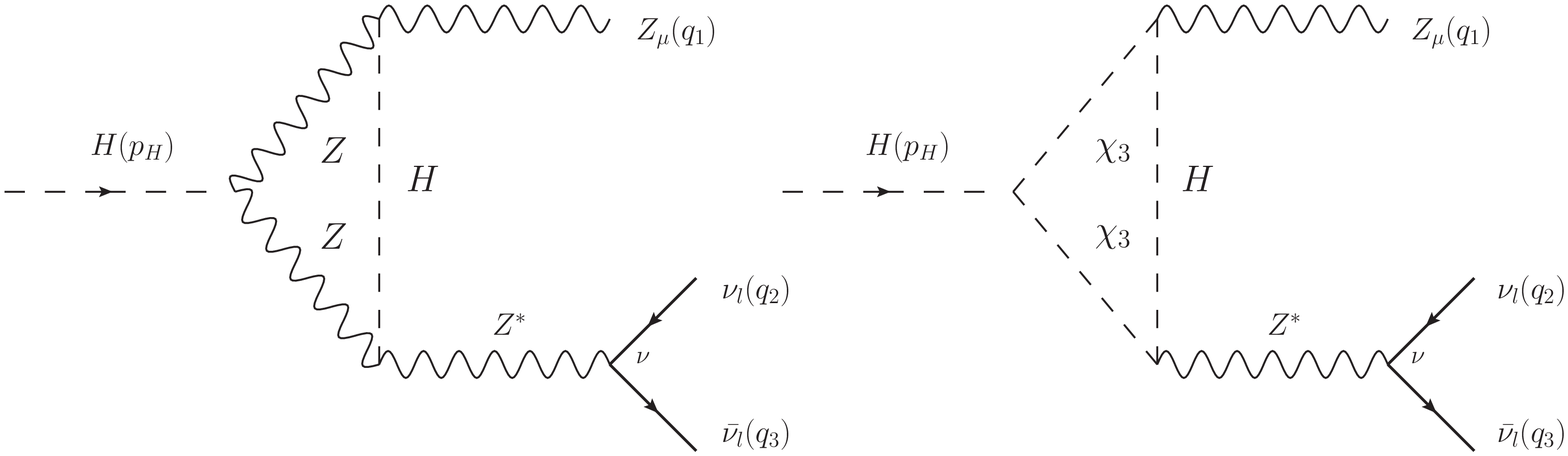}
\centering
\includegraphics[width=15.0cm, height=3.2cm]
{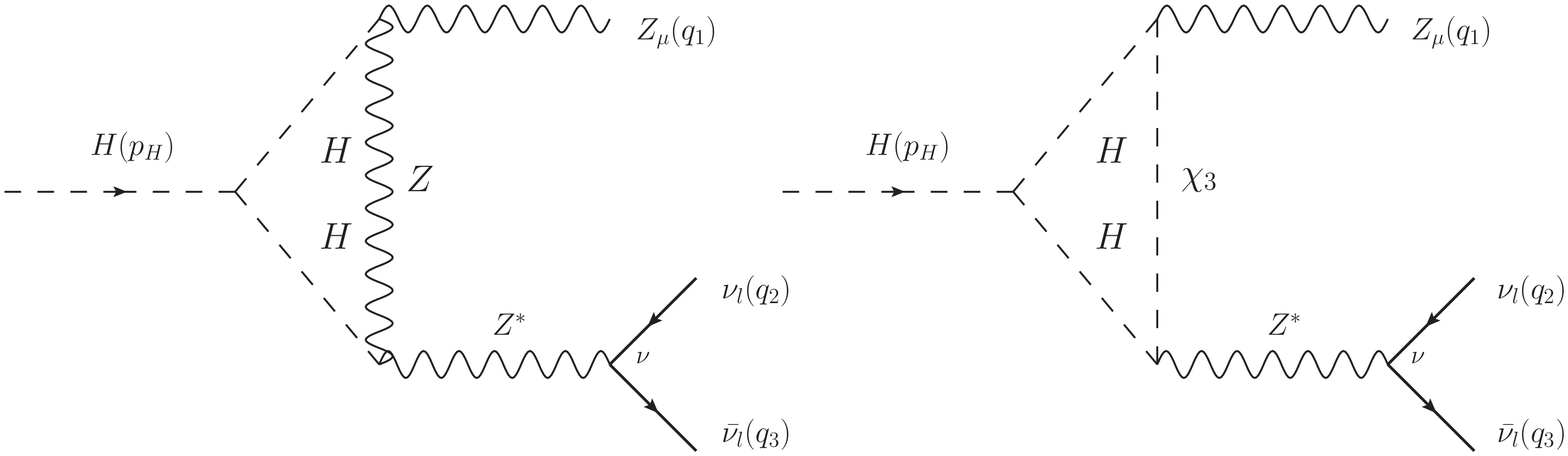}
\centering
\includegraphics[width=15.0cm, height=3.0cm]
{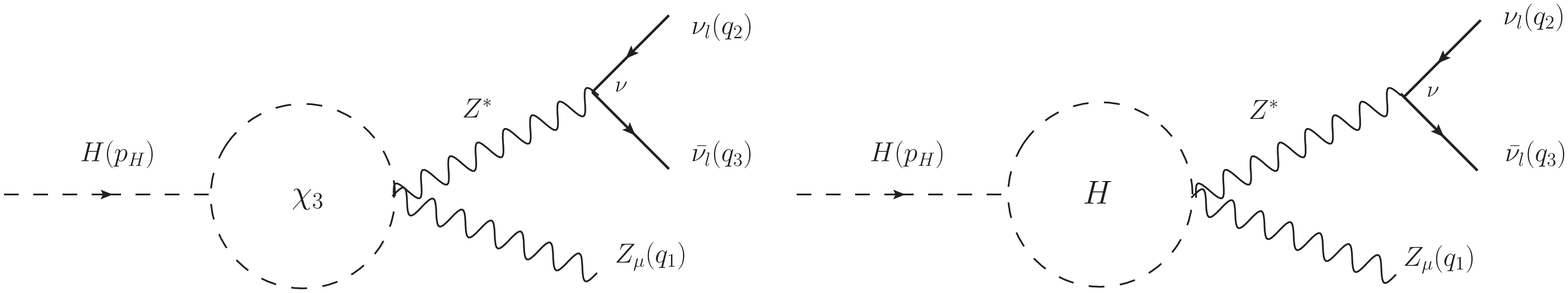}
\centering
\includegraphics[width=15.0cm, height=3.2cm]
{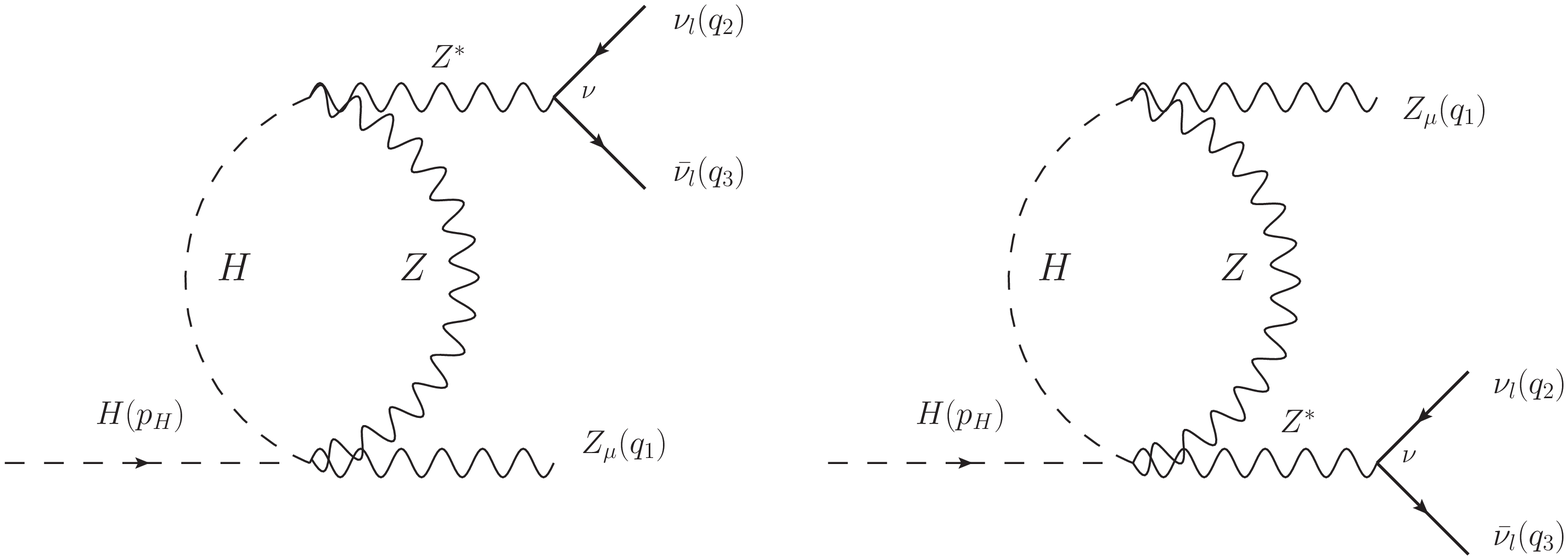}
\centering
\includegraphics[width=15.0cm, height=3.2cm]
{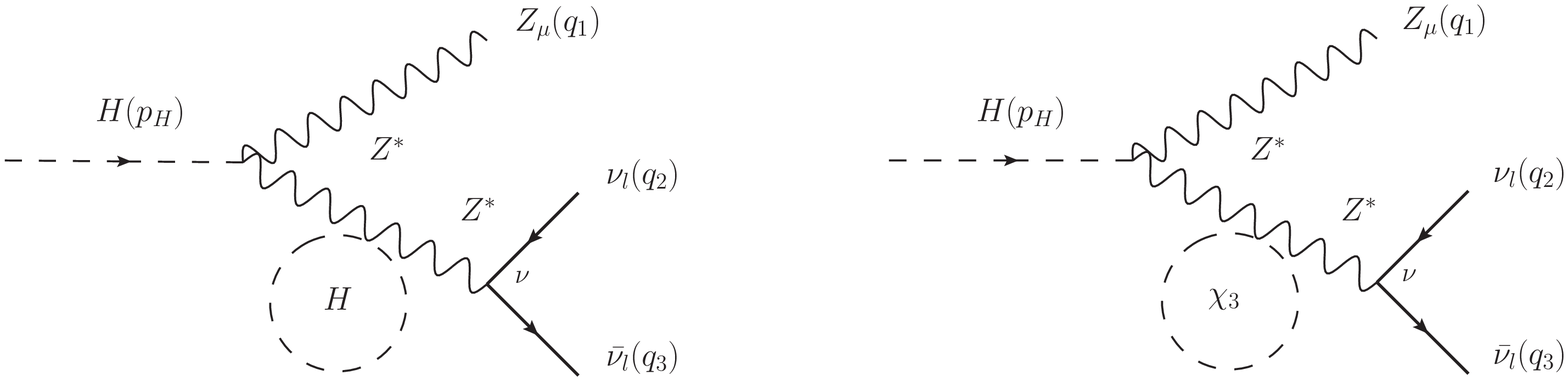}
\centering
\includegraphics[width=15.0cm, height=3.2cm]
{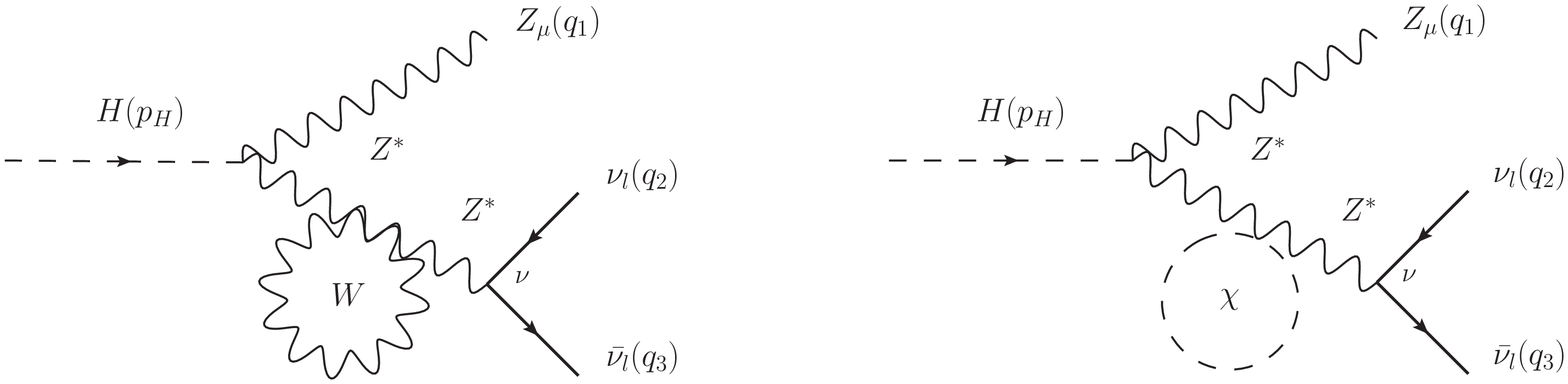}
\centering
\caption{Group $G_2$: All $Z$-pole Feynman
diagrams contributing to the decay process.
We note that $\chi_3$ is Nambu-Goldstone boson.}
\end{figure}
%%%%%%%%%%%%%%%%%%%%%%%%%%%%%%
\begin{figure}[ht]
\centering
\includegraphics[width=15.0cm, height=3.2cm]
{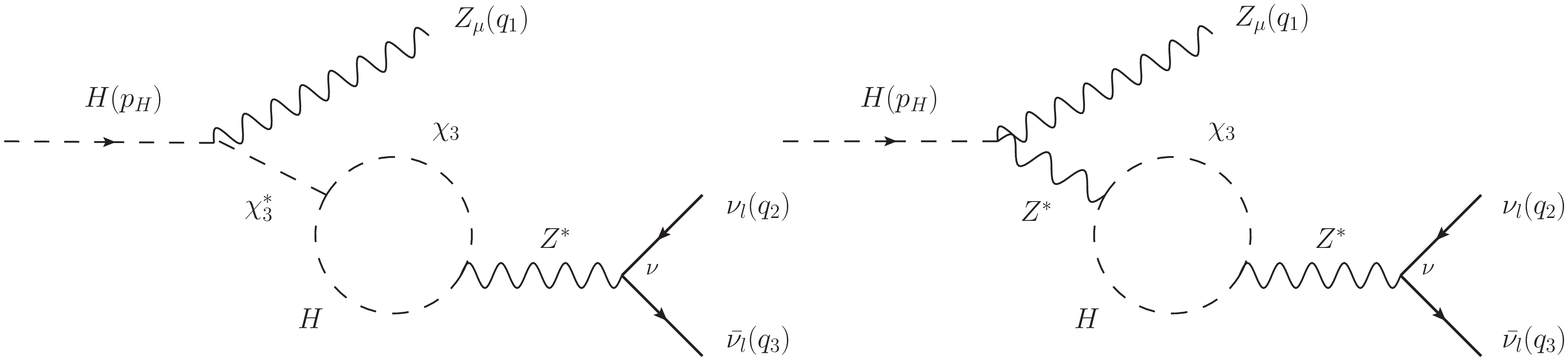}
\centering
\includegraphics[width=15.0cm, height=3.2cm]
{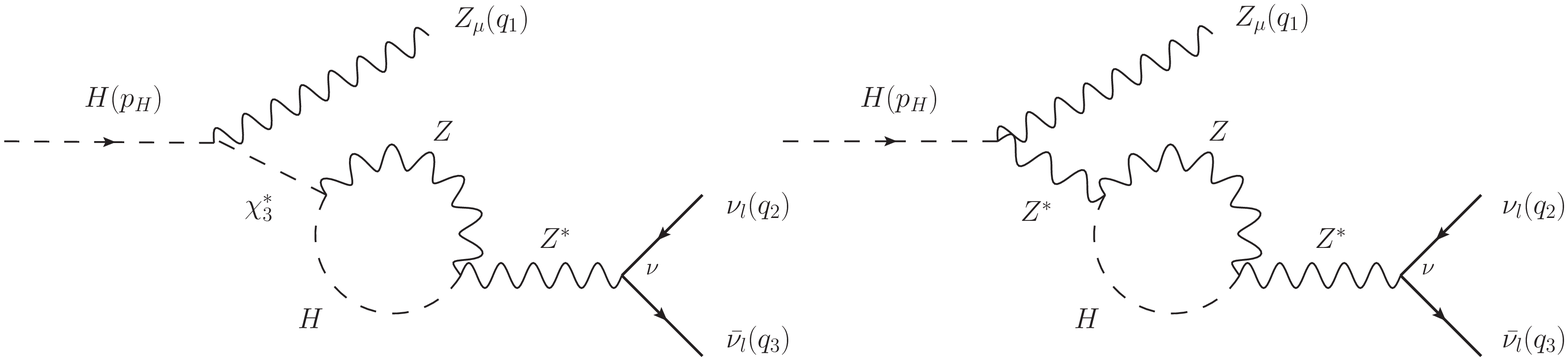}
\centering
\includegraphics[width=15.0cm, height=3.2cm]
{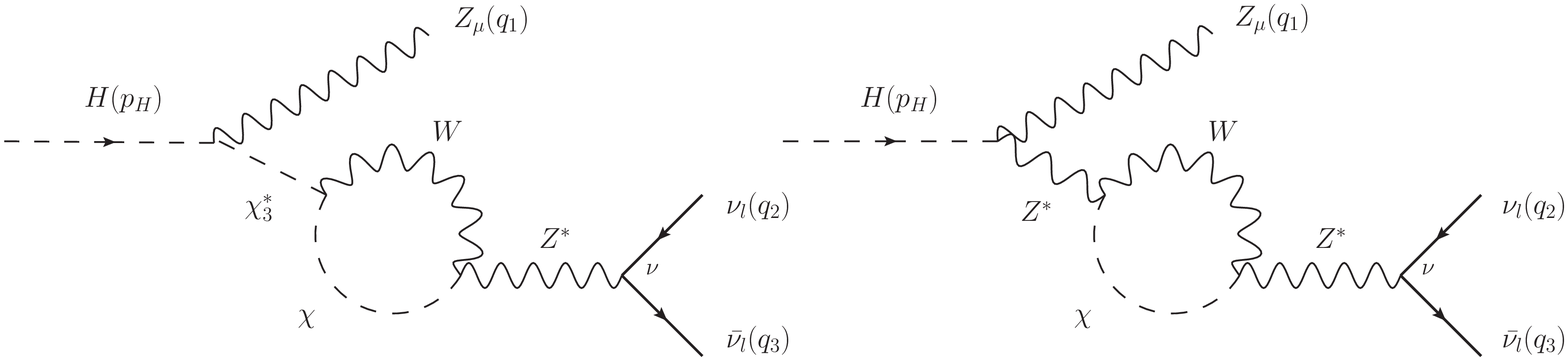}
\centering
\includegraphics[width=15.0cm, height=3.2cm]
{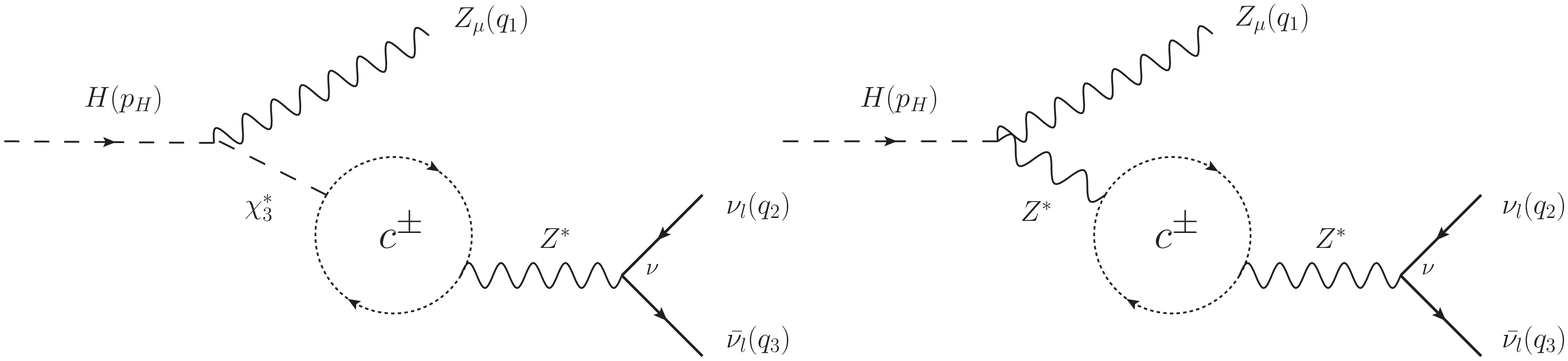}
\centering
\includegraphics[width=15.0cm, height=3.2cm]
{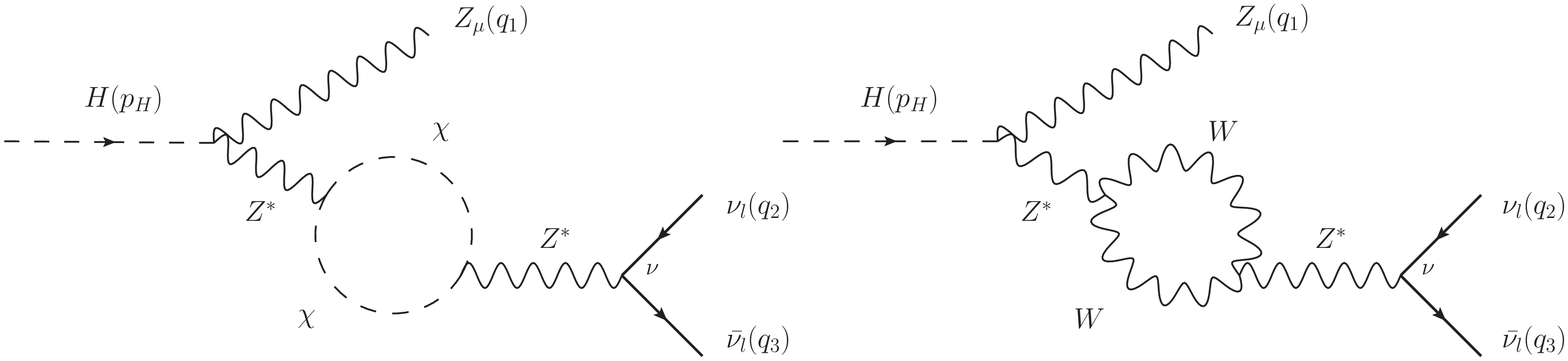}
\centering
\caption{Group $G_2$: All $Z$-pole Feynman
diagrams contributing to the decay process.
We note that $\chi^{\pm}$ and $c^{\pm}$
are Nambu-Goldstone bosons and 
ghost particles, respectively.}
\end{figure}

\begin{figure}[ht]
\centering
\includegraphics[width=15.0cm, height=3.2cm]
{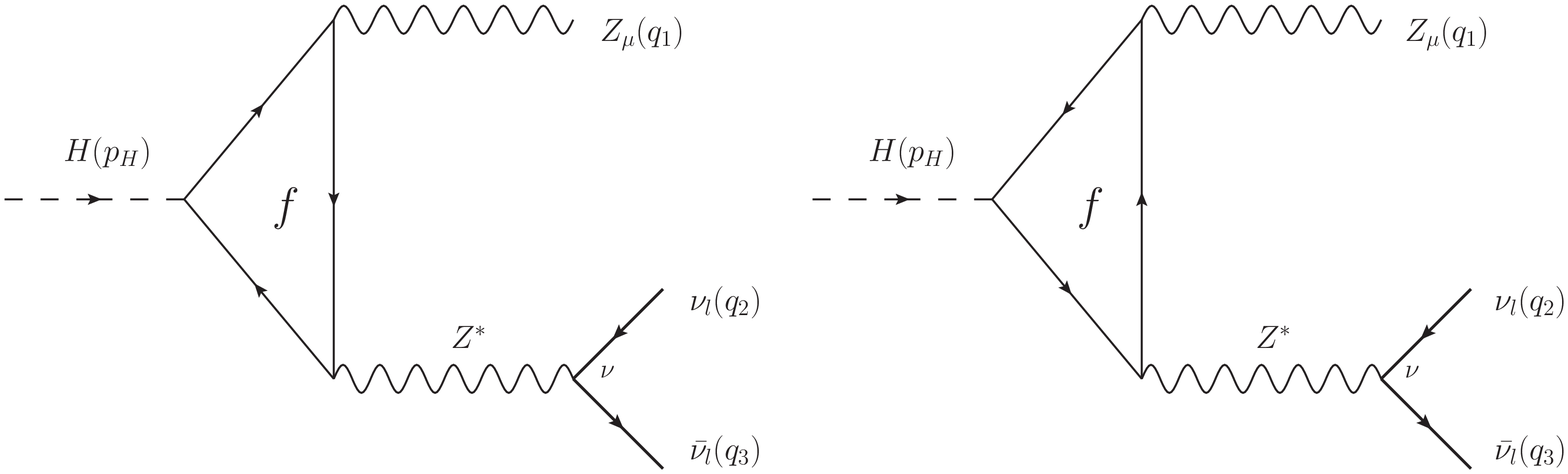}
\centering
\includegraphics[width=15.0cm, height=3.2cm]
{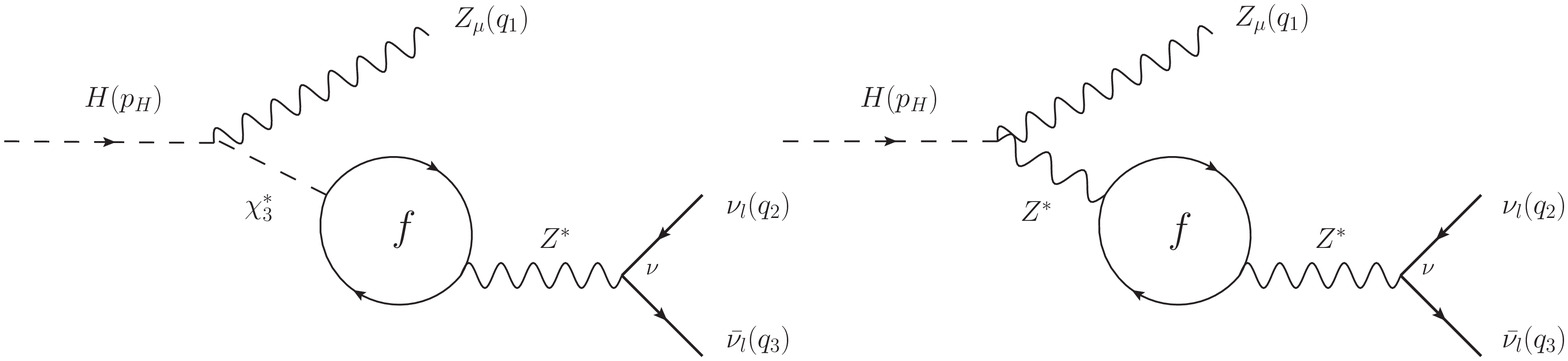}
\centering
\caption{Group $G_2$: All $Z$-pole Feynman
diagrams contributing to the decay process.
Here, $\chi_3$ is Nambu-Goldstone boson.
}
\end{figure}
%%%%%%%%%%%%%%%%%%%%%%%%%%%%
\begin{figure}[ht]
\centering
\includegraphics[width=15.0cm, height=4.0cm]
{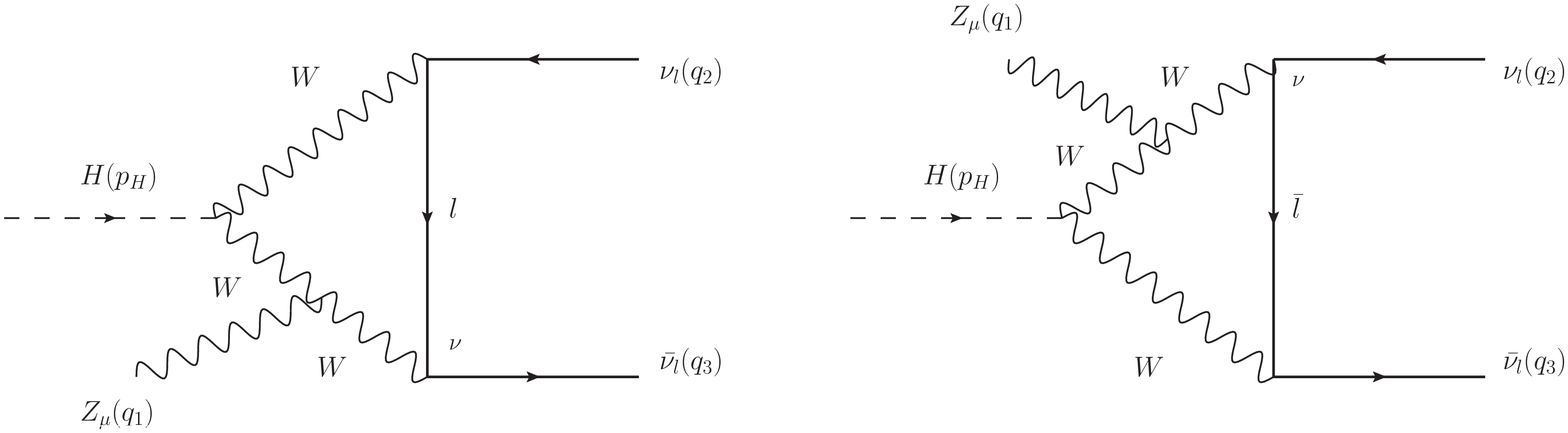}
\centering
\includegraphics[width=15.0cm, height=4.0cm]
{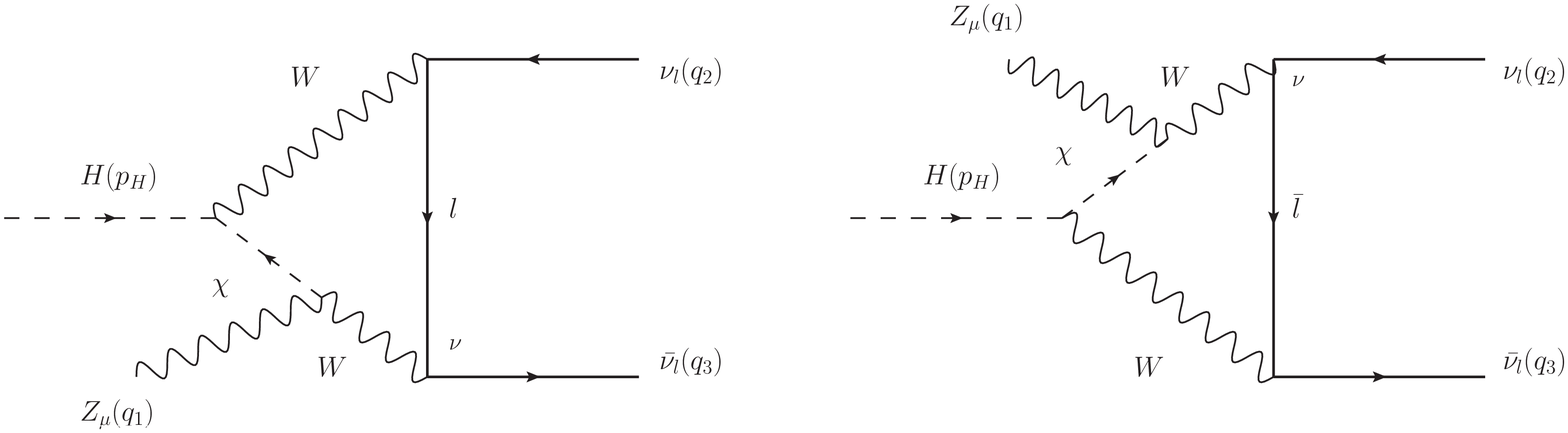}
\centering
\includegraphics[width=15.0cm, height=8.0cm]
{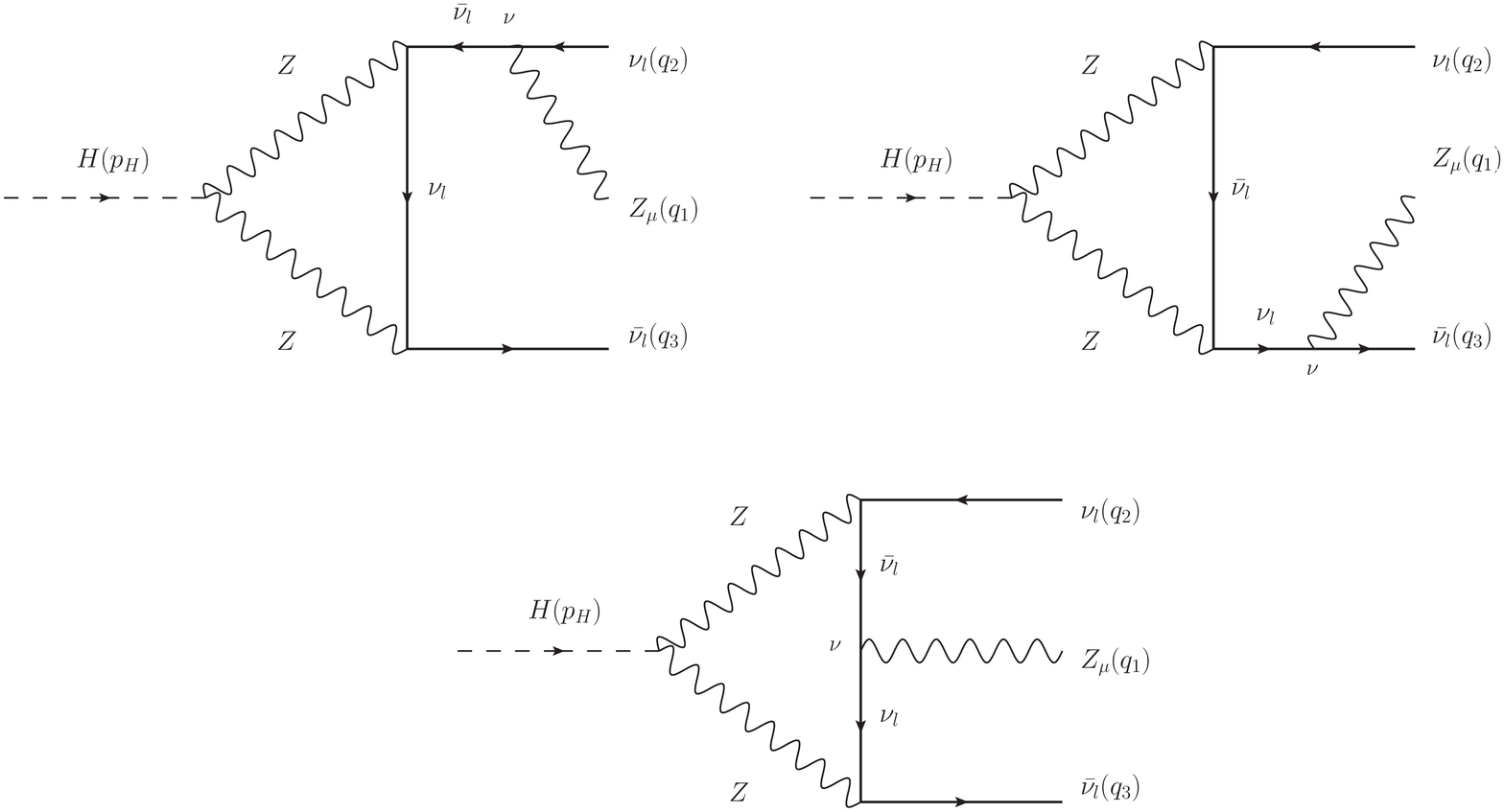}
\centering
\includegraphics[width=15.0cm, height=8.0cm]
{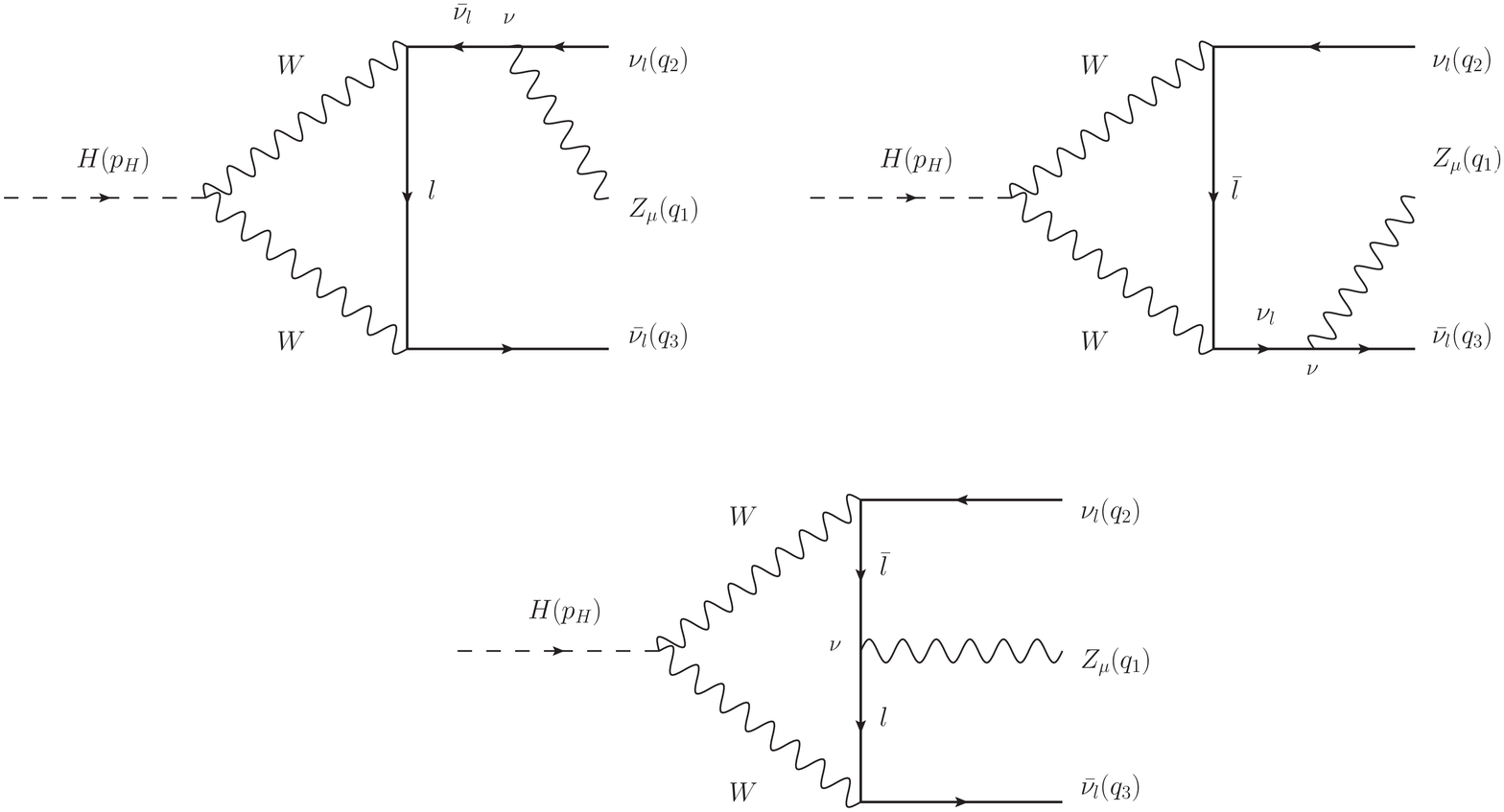}
\centering
\caption{Group $G_3$: All non $Z$-pole Feynman
diagrams contributing to the decay process.
Here $\chi^{\pm}$ 
are Nambu-Goldstone bosons.}
\end{figure}
%%%%%%%%%%%%%%%%%%%%%%%%%%%%%%
\begin{figure}[ht]
\centering
\includegraphics[width=16.0cm, height=10.0cm]
{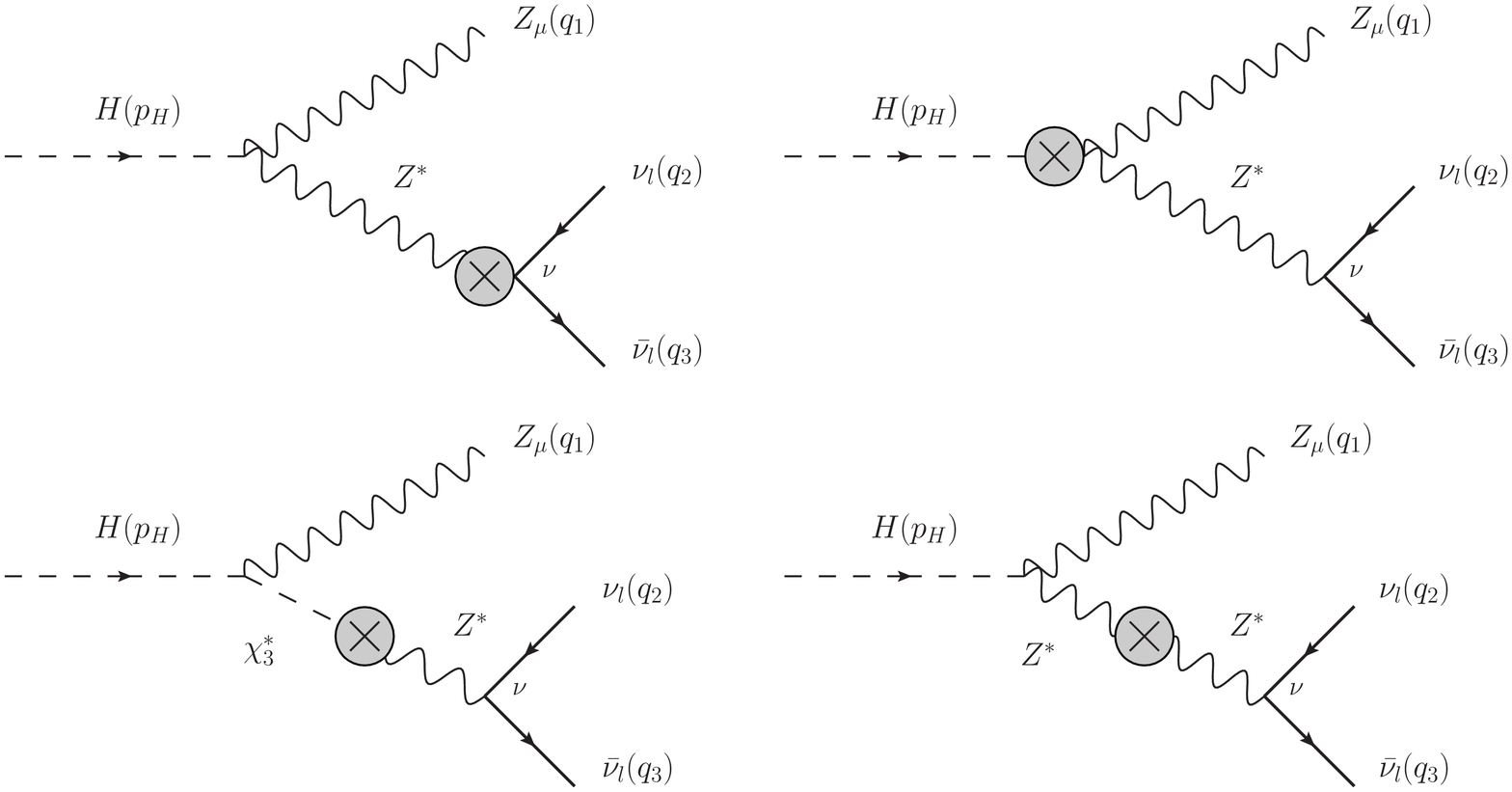}
\centering
\caption{Group $G_4$: All counterterm
Feynman
diagrams contributing to the decay process.}
\end{figure}

%%%%%%%%%%%%%%%%%%%%%%%%%%%%%%%%%%%%%%%%%%
%%%%%%
%%%%%%%%%%%%%%%%%%%%%%%%%%%
\end{document}